\documentstyle[12pt,epsf]{article}

\def \sect #1 {\setcounter{equation} 0\section{#1}}
\def \be  {\begin{equation}}
\def \ee  {\end{equation}}
\def \ba  {\begin{eqnarray}}
\def \ea  {\end{eqnarray}}
\def \baa {\begin{eqnarray*}}
\def \eaa {\end{eqnarray*}}
\def \bb  {\begin{thebibliography}}
\def \eb  {\end{thebibliography}}

\def \lab #1 {\label{#1}}

\def \fracs #1#2 {\mbox{\small $\frac{#1}{#2}$}}

\def \bin #1#2 {{\left({#1}\atop{#2}\right)}}
\def \as {\relax\ifmmode\alpha_s\else{$\alpha_s${ }}\fi}

\def \al #1 {\frac {\as({#1})}{\pi} }
\def \ds #1 {\ooalign{$\hfil/\hfil$\crcr$#1$}}

\newcommand \eh {\hat{\eta}}

\newcommand \bit {\begin{itemize}}
\newcommand \eit {\end{itemize}}

\newcommand \fmi {{\phi}_{min}}
\newcommand \fma {{\phi}_{max}}

\def \O {\Omega}

\def\hepph  #1 {{\tt hep-ph/#1}}

\begin{document}

\begin{flushright}
YITP-SB-01-35\\
BNL-HET-01/23\\
Revised\\
\today
\end{flushright}

\vspace*{30mm}

\begin{center}
{\LARGE \bf Energy Flow in Interjet Radiation}

\par\vspace*{20mm}\par

{\large \bf
Carola F.\ Berger$^a$, Tibor K\'ucs$^a$ and
George Sterman$^{a,b}$}

\bigskip

{\em $^a$C.N.\ Yang Institute for Theoretical Physics,
SUNY Stony Brook\\
Stony Brook, New York 11794 -- 3840, U.S.A.}

\bigskip

{\em $^b$Physics Department, Brookhaven National Laboratory\\
Upton, New York 11973, U.S.A.}

\end{center}
\vspace*{15mm}

\begin{abstract}
     We study the distribution of transverse energy, $Q_\O$,
radiated into an arbitrary interjet angular region, $\O$, in
high-$p_T$ two-jet events.
Using an approximation that
emphasizes radiation directly from the partons that undergo the
hard scattering,
we find a distribution that can be
extrapolated smoothly  to $Q_\O=\Lambda_{\rm QCD}$,
where it vanishes.  This method, which we apply numerically in a  
valence quark approximation,
provides a class of predictions on transverse energy radiated
between jets, as a function of jet energy and rapidity,
and of the choice of the region $\O$ in which the energy is measured.
We discuss the relation of our approximation
to the radiation from unobserved partons of intermediate
energy, whose importance was identified by Dasgupta and Salam.
\end{abstract}

\section{Introduction}

Jet events at hadron colliders encode
QCD time evolution over a large range of length
scales, from the momentum transfer that produced
the jets, to the scale of hadronization.
They thus afford a window to the
transition of quantum chromodynamics from
short to long distances, in addition
to  being our most direct probe of new
physics at short distances.
In this paper, we will study transverse energy flow into specified angular
regions between high-$p_T$ jets.  We will
find a variety of perturbative predictions,
in which soft
radiation is sensitive to color and flavor flow
at short distances.
The flow in soft
radiation also has an impact on the comparison
of measured cross sections in a hadronic
environment to perturbative calculations.  These measurements rely on our
determination of the energies of  jets in the presence of
an ``underlying event", consisting of particles not
directly associated with jet production.
Because jet cross sections fall so rapidly with momentum transfer,
they are sensitive to relatively small shifts
in energy \cite{Soper,Mangano}, even at the highest $p_T$.
This is the case, even though the effects of
such shifts are power-suppressed in $p_T$,
compared to finite-order perturbative calculations.
An understanding of the underlying event requires
good control over perturbative bremsstrahlung associated
with the hard scattering itself, so that the two effects
may be separated.

The influence of the underlying event on
jet energy measurements
is a complex issue \cite{Webber,HuTa,Field}.
In the absence of further information, it is tempting
to think of the particle production underlying the jets
as similar to that in a minimum bias event, one in which there is no
distinguishing hard scattering.
Energy flow between jets in the final
state, however, may result from multiple parton
interactions \cite{cdfds,jp}, which may or may not be the
same in jet and minimum bias events, as well as
from perturbative QCD bremsstrahlung
associated with the hard scattering itself  \cite{ua2,cdf91}.
Marchesini and Webber \cite{mw88} argued
that it should be possible to disentangle
the underlying event from the bremsstrahlung of the
hard scattering by studying the flow of transverse energy
away from the jet direction.  They pointed out
that the average transverse energy of radiation
into specified angular regions is infrared safe.
We will develop here a further technique to compute
the {\it distribution} in transverse energy of this radiation.

The other motivation for this study of interjet
radiation is the light
it may shed on short-distance color and flavor flow,
and on the dynamics of hadronization.
The relation of the angular dependence of interjet
particle  multiplicity to color flow in the hard
scattering has been well-established in
$\rm e^+e^-$ annihilation \cite{coco}.    Its use
in the  analysis of jet events in hadronic
scattering, to distinguish
new physics signals from QCD background, was
explored in \cite{EKS,KhSt}, and  the method was
verified experimentally in \cite{d0jetZ},
through the analysis of events involving jets and
W bosons.  Here we shall follow  \cite{mw88}, and investigate
transverse energy flow rather than  multiplicity, but the spirit of
the approach is much the same.

The discussion below is also
closely related to
the treatment of rapidity gap
(``color singlet exchange")
events in Refs.\ \cite{StOd,Oderda},
for jets at high $p_T$ and large rapidity
separations.  In this case, we will study
relatively central pairs of jets, with the high-$p_T$ tail
of the inclusive jet cross section in mind.
Correspondingly, we restrict ourselves to a valence
approximation.

We will calculate the distribution for the flow of transverse
energy, $Q_\O$, radiated into a specific detector region, $\Omega$, away
from
a set of observed jets and from the beam directions.
   This is what we mean by interjet
radiation.\footnote{In fact, our analysis can be applied to a
class of kinematic quantities, including the energy itself.
We choose to work with the transverse energy simply
because of its boost invariance.}
The distribution in $Q_\O$ is perturbatively calculable
as a factorized cross section,
although for $Q_\O\ll Q$, with $Q$ the hard momentum transfer,
the perturbative expansion will in
general include contributions of order $(1/Q_\O)\, \alpha_s^n\, 
\ln^{n-1}(Q_\O/Q)$.
We shall resum such terms in a leading logarithmic approximation,
in the ratio $p_T/Q_\O$, with $p_T$ the transverse momentum
of the final-state jets.  We shall assume  that there are no additional jets
of comparable energy in the final state.  More specifically,
we sum inclusively on energy only within cones around the
beam and outgoing jet directions.
We denote by $\bar \O$ the union of all directions outside of
these cones and outside of $\O$.
To enforce our two-jet
criterion, we allow a maximum total energy $p_T\gg M\gg Q_\O$ into $\bar \O$.
More generally, the parameter $M$ can be thought of as
shorthand for an angular-dependent cutoff on energy flow
into $\bar \O$.

As shown recently by Dasgupta and Salam \cite{DasSal01,DasSal02} in studies
of $\rm e^+e^-$ annihilation, partons emitted into $\bar \O$
influence energy flow into $\O$ through secondary radiation,
producing terms that behave as $(1/Q_\O)\, \alpha_s^n\ln^{n-1}(M/Q_\O)$,
for $n\ge 2$.  They call these contributions to the
cross section ``non-global" logarithms.
We shall see how they arise in our cross sections,
but we shall simplify our discussion by neglecting
logarithmic corrections in $M/Q_\O$.
We comment on the implications
of this approximation below.  The interplay of
the definition of jet events and non-global logarithms
will be the subject of forthcoming work.

Our
analysis is made possible by the incoherence
of wide-angle soft gluon radiation from the internal
evolution of jets,
a property of QCD which plays an important role
both in proofs of factorization \cite{fact}, and in
the study of jet properties through the
coherent branching formalism \cite{coco,cobr}.
We will see how non-global logarithms
arise in this context from the radiation of gluons up to the maximum
allowed energy $M$ in region $\bar \O$.  These hard gluons themselves
then act as color sources for radiation into $\O$.  Nevertheless,
at leading logarithm, because they are emitted at wide
angles, the eikonal approximation for radiation from an underlying 
two-to-two hard
scattering remains.

We show below
  how interjet radiation,
although independent of the details
of the jets, depends upon both
flavor and color exchanges
at the hard scattering, and we will
show how to derive the distribution
of events in terms of $Q_\Omega$ for each combination.
The choice of $\Omega$ is free, so long as it does
not include the observed jets.
In this paper, we illustrate our method
with the relatively simple choice
of a region whose area is of order unity in
the $\eta-\phi$ space of rapidity and azimuthal angle, outside of
the beam-jet scattering plane.

In the following section, we discuss the class of interjet energy flow
cross sections that we will study,
starting from their collinear-factorized form.
We describe the refactorization of hard-scattering
functions into fully short-distance functions,
dependent only on scales at the
level of the large momentum transfer, $Q^2\sim p_T^2$, and a set of
``soft'', but perturbatively
calculable, functions, which describe evolution
at scales intermediate between $Q^2$ and
$Q_\O^2$.  To apply perturbation theory, of
course, we must assume that $Q_\O$ is
large compared to the scale at which
the running coupling becomes strong: $Q_\O^2\gg \Lambda_{\rm QCD}^2$.
We show how this refactorization leads to
renormalization group equations for the soft
functions, and hence for their $Q_\O$-dependence.
These are
matrix equations describing the mixing of color \cite{KOS2}.
      Section 3 deals with
the determination of the anomalous dimension matrices
that control this evolution.  We explicitly
calculate the matrices for initial-state quarks
and antiquarks.  We show how, at
leading logarithm, their calculation reduces to
the evaluation of integrals over the phase
space of radiation into $\Omega$.  In Section
4, we study specific distributions for
scales relevant to Tevatron
   high-$p_T$ two-jet events.
We close with a brief discussion of our results,
and prospects for future progress.

\section{Factorization and Refactorization}

\subsection{Dijets and interjet observables}

We will discuss inclusive cross sections involving dijet production
in hadronic collisions:
\be
A(p_A) + B(p_B) \rightarrow J_1(p_1) + J_2(p_2) + R_\O(Q_\O) + X_{\bar \O}\, ,
\label{2jetX}
\ee
where the jets $J_i$ have high transverse momenta.
In principle, the jet axes may be identified
by any jet-finding convention,
but we shall always define the jet energy by
flow into some angular region, a ``cone".
In Eq.\ (\ref{2jetX}),  $R_\O$ denotes radiation
into a region $\O$ outside these jet cones.
The quantity $Q_\Omega$
is a kinematic measure of the radiation into $\Omega$.
We may, as in Refs.\ \cite{StOd,Oderda}, take $Q_\Omega$ to
be the energy of particles in  $\Omega$, but we may also take it to be
the
transverse energy.
As above, we refer to the region between $\O$ and these cones by $\bar \O$.
The jets are
characterized by their transverse
momenta and pseudorapidity,
$\eta = \ln(\cot(\theta/2))$,
where $\theta$ is the angle
with respect to the
beam axis.
We will consider two-jet final states for which
\be
|\vec p_{T1}|\sim |\vec p_{T2}| \gg Q_\Omega\gg \Lambda_{\rm QCD}\, .
\label{ptgeqo}
\ee
We sum inclusively over the radiation $X_{\bar{\O}}$ in
$\bar \O$, up to a cutoff, $M$ on its total energy
in the partonic center-of-mass (c.m.) frame,
with $p_T\gg M \gg Q_\O$.
Thus, the complete interjet radiation
into $\bar \O$ and $\O$, although energetic enough to
be treated in perturbation theory, is
assumed to be soft enough
that we need not consider the recoil of the high-$p_T$
jets.

Collinear factorization theorems \cite{fact}
ensure that we may write
any such cross section at fixed $Q_\O$
as a sum of convolutions of parton
distribution functions $\phi$ (PDFs)
that incorporate long-distance dynamics,
with hard-scattering functions, $\omega$,
         that summarize short-distance dynamics,
\ba
\frac{d\sigma_{AB}}{d\vec p_1\, dQ_\O}
&=&
\sum_{f_A,f_B} \int dx_A\, dx_B\;
{\phi}_{f_A/A}\left(x_A, \mu_F\right) {\phi}_{f_B/B}
\left(x_B,\mu_F\right)\nonumber\\
&\ & \hspace{15mm} \times\,
\omega_{f_Af_B}\left(x_Ap_A,x_Bp_B,\vec
p_1,M,Q_\O,\mu_F,\alpha_s(\mu_F)\right)\, ,
\label{cofact}
\ea
with factorization scale $\mu_F$.
The sum is over parton types, $f_A,f_B$.
Corrections to
this relation begin in general with powers
of $\Lambda^2_{\rm QCD}/Q^2_\O$, due to  multiple
scattering of partons -- part of the underlying event
referred to above.   Our analysis below, which
begins with Eq.\ (\ref{cofact}), is thus accurate
up to such corrections, and requires the
conditions (\ref{ptgeqo}).
We take the renormalization scale
equal to the factorization scale. Because in this paper
we will emphasize ratios of cross sections rather than
their size, fixing $\mu_{\rm ren} = \mu_F=p_T$ is
meant to be taken as an indication of the order
of these scales, rather than the promotion
of a specific choice.

\subsection{Refactorization of the partonic cross section}

As stated above, we define the energies of high-$p_T$
jets by the energy flowing into specified cones, while the
region $\O$ in which we measure $Q_\O$ is outside
these cones.
Soft gluon emission outside the angular extent of the jets
decouples from the kinematics of the hard scattering,
and from the internal evolution of the jets \cite{coco,fact,cobr}.
As a result, we may express $\omega_{f_Af_B}$ in Eq.\ (\ref{cofact})
as a sum of terms, each characterized by a definite
number of jets produced at the hard scattering.
To lowest order in $\alpha_s$, only two jets are possible.
Thus, at leading order, the production
of the high-$p_T$ jets is given by the set of Born-level
$2\rightarrow 2$ partonic processes, which we label f,
\be
{\rm f}: \hspace{5mm} f_A+ f_B\rightarrow f_1+ f_2\, .
\label{pprocess}
\ee
We distinguish $q \bar{q} \rightarrow q \bar{q}$ ($f_1 = q$) from
$q \bar{q} \rightarrow \bar{q} q$  ($f_1 = \bar{q}$).
We may now write the single-jet
inclusive cross section at fixed $Q_\O$ as
\ba
\frac{d\sigma_{AB}}{d\eta dp_T dQ_\O}
= \sum_{\rm f} \int dx_Adx_B \;
{\phi}_{f_A/A}\left(x_A, \mu_F\right) {\phi}_{f_B/B}
\left(x_B,\mu_F\right)\, \delta\left( p_T-{\sqrt{\hat s}\over
2\cosh\hat\eta}\right)\,
\frac{d \hat{\sigma}^{\rm (f)}}
{d\hat\eta  dQ_\O}\, ,
\label{hatsigfact}
\ea
with $\hat\eta=\eta-(1/2)\ln(x_A/x_B)$ the jet
rapidity in the partonic center-of-mass, and
$\hat s=x_Ax_Bs$.
  We neglect the effects
of recoil of the observed jet against relatively soft
radiation.
The short-distance
function ${d \hat{\sigma}^{\rm (f)}}/{d\hat \eta   dQ_\O}$
is an expansion in $\alpha_s(p_T)$
that begins with the lowest order (LO) cross section,
\be
\frac{d \hat\sigma^{\rm (f)}}{d\hat\eta   dQ_\O} =
\frac{d\hat\sigma^{\rm (f,LO)}}{d\hat\eta }\,
\delta(Q_\O)+\dots\, ,
\label{hatsigexpand}
\ee
but which includes
potentially large corrections associated with logarithms
of $Q_\O/p_T$.
In the two-jet approximation described above,
we work at leading order in the hard scattering,
with the factorization
scale $\mu_F = p_T$.
The  sum runs over all possible partonic subprocesses f,
defined as in (\ref{pprocess}).
We emphasize that corrections to Eq.\ (\ref{hatsigfact})
begin at ${\cal O}(\alpha_s(p_T))$, due to processes
with more than two energetic jets in the final state.
Taking these into account would require generalizing
the set of flavor flows f in Eq.\ (\ref{pprocess})
to $2\rightarrow 3$ and beyond.  Such a development
will be informative, but we shall not pursue it
here.

The hard-scattering function
for the subprocess f may now be refactorized \cite{KOS2,KOS1}, to
exhibit the interrelation between the partonic
hard scattering and soft radiation
outside all jet cones.
For this purpose, it is convenient to treat the partonic
and hadronic cross sections integrated over
the transverse
energy radiated into region $\O$, up to a fixed value $Q_\O$,
\ba
{d\hat \sigma^{\rm
(f)}\left(p_T,\hat\eta,\mu_F,M,Q_\O,\alpha_s(\mu_F)\right)
           \over d\hat\eta  } &=&\int_0^{Q_\O} dQ_\O'
{d\hat \sigma^{\rm
(f)}\left(p_T,\hat\eta,\mu_F,M,Q'_\O,\alpha_s(\mu_F)\right)
           \over d\hat\eta   dQ'_\O} \nonumber\\
{d \sigma_{AB}\left(p_T,\eta,\mu_F,M,Q_\O,\alpha_s(\mu_F)\right)
           \over d\eta  d p_T }
&=& \sum_{\rm f} \int dx_Adx_B\; \phi_{f_A/A}(x_A,\mu_F)\,
\phi_{f_B/B}(x_B,\mu_F)
 \label{inclusive}\\
&\ & \hspace{10mm} \times \,\delta\left( p_T-{\sqrt{\hat s}\over
2\cosh\hat\eta}\right)\,
{d\hat \sigma^{\rm
(f)}\left(p_T,\hat\eta,\mu_F,M,Q_\O,\alpha_s(\mu_F)\right)
           \over d\hat\eta   } \, . \nonumber
\ea
As we shall show,
the relevant refactorization of the
hard scattering function of Eq.\ (\ref{inclusive}) is
\be
\frac{{d\hat{\sigma}}^{\rm (f)} \left(p_T, \hat
\eta,\mu_F,M,Q_\O,\alpha_s(\mu_F)\right)}
{d\hat\eta   } =
\sum_{L,I}\;
         H^{\rm (f)}_{IL}\left(p_T,\hat\eta, \mu, \mu_F, 
{\alpha}_s(\mu) \right)
S^{\rm (f)}_{LI}\left(\hat\eta,\Omega,
\frac{M}{\mu},\frac{Q_\O}{\mu},\alpha_s(\mu)\right)\,  ,
\label{refact}
\ee
where $H_{IL}^{\rm (f)}$ is a matrix of short-distance functions,
independent
of $Q_\O$, while all of the $Q_\O$-dependence is factored
into dimensionless soft functions $S_{LI}^{\rm (f)}$.
Here $\mu$ is a new refactorization scale,
chosen between $p_T$ and $Q_\O$, with
dynamics on scales below $\mu$ incorporated into $S^{\rm (f)}$.
The soft function is independent of $\mu_F$, which
separates $d \hat{\sigma}/d\hat{\eta}$ from the
parton distribution functions.
The functions $S_{LI}^{\rm (f)}$, which summarize
dynamics of partons with energies at the scale $Q_\O$, depend
in general on the scalar products
$\beta_i\cdot \beta_j$ between the
lightlike 4-velocities of
the incoming and outgoing energetic partons in the hard scattering,
which radiate soft gluons. If we fix the scattering plane
such that jet 1 is located at $\phi = 0$, the kinematics of
two-to-two scattering reduces this
dependence to a single variable, which we may
take as $\hat\eta$, the
rapidity of parton 1 (see Eq.\ (\ref{pprocess}))
in the partonic c.m.\ frame.
Note that $S^{\rm (f)}_{LI}$ is also a function
of the maximum energy, $M$, radiated into the
complementary interjet region, $\bar\O$.

In Eq.\ (\ref{refact}), the hard-scattering functions
$H_{IL}$ begin at order
$\alpha_s^2$,
while the soft functions $S_{LI}$ begin
         at zeroth order.
To leading logarithm, we need only these lowest-order
contributions, denoted below as $H^{\rm (f,\; 1)}$ and
$S^{({\rm f},\; 0)}$, respectively.
The combination of these approximations is
related to the lowest-order cross section by Eq.\ (\ref{hatsigexpand}),
\ba
\frac{d\hat\sigma^{\rm (f,LO)}}{d\hat\eta  }\,
=
\sum_{L,I}\;
         H^{\rm (f,\; 1)}_{IL}\left(p_T,\hat\eta, {\alpha}_s(\mu_F)\right)
S^{\rm (f,\; 0)}_{LI}\, .
\label{lo}
\ea
Recalling that we expect corrections
to Eq.\ (\ref{hatsigfact}) to begin at
NLO in $\alpha_s(p_T)$
for inclusive jet production,
Eq.\ (\ref{refact}) should hold to a corresponding accuracy
in the hard-scattering function $H_{IL}$.
Additional corrections to the refactorization (\ref{refact})
may be due to multiple scattering.  These
corrections are suppressed by
powers of ${\Lambda}^2_{\rm QCD}/Q^2_\O$.
At large $Q_\O$, therefore, bremsstrahlung dominates
the inclusive cross section.   At low $Q_\O$,
the NLO differential
cross section for bremsstrahlung
diverges as $1/Q_\O$.
As we shall see, however, the resummed \emph{integrated} cross section,
Eq. (\ref{inclusive}),
is better behaved, and may be extrapolated
smoothly as far down as $Q_\O=\Lambda_{\rm QCD}$,
where it vanishes.  In these
terms, the perturbative component of interjet radiation,
which dominates for hard radiation, joins smoothly
to the region of low radiation, and may thus be
meaningfully compared to data \cite{mw88}.

\subsection{The soft function}

The proof of Eq.\ (\ref{refact}) at the
desired accuracy, leading logarithm in $Q_\O$,
follows standard arguments in factorization \cite{fact,gsTasi95}.
An essential ingredient in proofs of collinear factorization
is that wide-angle soft radiation decouples from
jet evolution \cite{CS81}.  Soft
radiation away from jet directions is equally well
described by radiation from a set of path-ordered
exponentials -- nonabelian phase operators
or Wilson lines -- that replace each of the partons
involved  in the hard scattering (four, in our case),
\ba
\Phi_{\beta}^{(f)}(\infty,0;x)
&=&
P\exp\left(-ig\int^{\infty}_{0}d{\lambda}\;
{\beta}{\cdot} A^{(f)} ({\lambda}{\beta}+x)\right)\, ,
\nonumber\\
\Phi_{\beta'}^{(f')}(0,-\infty;x)
&=&
P\exp\left(-ig\int_{-\infty}^{0}d{\lambda}\;
{\beta'}{\cdot} A^{(f')} ({\lambda}{\beta'}+x)\right)\, ,
\label{inoutPhidef}
\ea
where $P$ denotes path
ordering.  The first line describes an outgoing, and
the second an incoming, parton,
whose flavors and four-velocities are  $f$ and
$\beta$, and $f'$ and $\beta'$, respectively.
The vector potentials $A^{(f)}$
are in the color representation appropriate to flavor $f$,
and similarly for $A^{(f')}$.

Because wide-angle, soft
radiation is independent of the internal
jet evolution, products of
nonabelian phase operators,
linked at the hard scattering
by a tensor in the space of color indices
generate the same wide-angle radiation as the full jets.
The general form for these operators,
exhibiting their color indices, is
\ba
W_{I\, \{c_i\}}^{\rm (f)}(x) &=& \sum_{\{d_i\}}
\Phi_{\beta_2}^{(f_2)}(\infty,0;x)_{c_2,d_2}\;
\Phi_{\beta_1}^{(f_1)}(\infty,0;x)_{c_1,d_1}
\nonumber\\
&\ & \ \times \left(c_I^{\rm (f)}\right)_{d_2,d_1;d_A,d_B}\;
\Phi_{\beta_A}^{(f_A)}(0,-\infty;x)_{d_A,c_A}\;
\Phi_{\beta_B}^{(f_B)}(0,-\infty;x)_{d_B,c_B}
\, .
\label{Wdef}
\ea
The $c_I$ are color tensors in a convenient basis.
Examples will be given below.
The perturbative expansions
for these operators are in terms of standard eikonal
vertices and propagators, and have been given in
detail in Refs.\ \cite{KOS2,KOS1}.
      In these terms, we define
$S_{LI}^{\rm (f)}$ as
\footnote{This definition of $S^{\rm (f)}$ differs slightly from
that of Ref.\ {\protect\cite{StOd,KOS2,KOS1}}, in that it is
an integral over $Q_\O$, rather than differential in $Q_\O$.}
\ba
S^{\rm (f)}_{LI}\left(\hat\eta,\O,\frac{M}{\mu},
\frac{Q_\O}{\mu},\alpha_s(\mu)\right) &=& \int_0^{Q_\O} dQ'_\O\,
\sum_n \sum_{\{b_i\}}\, \delta(Q_\O^{(n)}-Q'_\O)\; 
\theta\left(M-E_{\bar \O}^{(n)}\right)
\nonumber\\
&\ & \hspace{5mm} \times \langle 0|\,
\bar T\left[\left(W^{\rm (f)}_L(0)\right)_{\{b_i\}}^\dagger\right]\,
|n\rangle
\langle n|\, T\left[W^{\rm (f)}_I(0)_{\{b_i\}}\right]\, |0\rangle\, .
\label{Sdef}
\ea
The sum is over all states $n$ whose
transverse energy flow into $\Omega$
is restricted to equal $Q'_\O$, and the energy into
$\bar \O$ is less than or equal to $M$.
The indices $L$ and $I$ refer to the color exchange at the
hard scattering between the partons in reaction f,
as built into the definitions of the $W$'s,
Eq.\ (\ref{Wdef}).   The matrix elements  in Eq.\ (\ref{Sdef})
require renormalization, and we may identify the
corresponding renormalization scale with the
refactorization scale of Eq.\ (\ref{refact}).

The $Q_\O$-dependence of the
matrix $S_{LI}^{\rm (f)}$ in Eq.\ (\ref{Sdef}) is  the same as in the
full
partonic cross section, up to corrections due to differences between
the jets and the nonabelian
phase operators.  Radiation within the jets, however, is treated
inclusively in the sum over final states in (\ref{Sdef}).
As observed above, radiation outside the jet cone is emitted coherently
by the entire jet \cite{coco,fact,cobr}.
Any logarithmic enhancements associated with (angular ordered)
radiation within the jets, or with radiation
collinear to the eikonal lines, are thus independent of $Q_\O$,
and cancel in the sum over states.  This cancellation
occurs at fixed momenta for all on-shell
particles within the jets \cite{CS81,S78}.
As a result, in the leading logarithmic
approximation, differences between the eikonal and exact
cross sections  are given by
$Q_\O$-independent  expansions in $\alpha_s(\mu)$,
which may be absorbed into the hard functions $H_{LI}$.
The $H$'s themselves are thus the coefficient
functions that match the effective eikonal operators
$W_I^{\rm (f)}$ to the full theory.  Therefore, Eq.\ (\ref{refact}),
with the soft function defined as in Eq.\ (\ref{Sdef}),
reproduces the $Q_\O$-dependence of the original
cross section, to leading logarithm, the level of the original
collinear factorization form restricted to a $2\rightarrow 2$
hard scattering, Eq.\ (\ref{hatsigfact}).

We shall see in the next subsection that
the color flows associated with $2\rightarrow 3$ and
beyond play an important role
in the analysis of the soft matrix $S_{LI}^{\rm (f)}$ itself, even at
leading logarithm \cite{DasSal01}.  These effects do not,
however, affect the basic refactorization of eikonal from
partonic degrees of freedom, Eq.\ (\ref{refact}).
In addition, as noted above, beyond leading logarithm the
sums over $I$ and $L$ in the factorized hard
scattering cross section, Eq.\ (\ref{refact}), must also be generalized
to include the possibility of $2\rightarrow 3$
and other scattering processes, in which
more lines emerge from the short-distance
processes.
To include these higher-order processes,
we would generalize the
$2\rightarrow 2$ eikonal
cross sections of $S^{\rm (f)}_{LI}$ of Eq.\ (\ref{Sdef})
to $2\rightarrow 3$,  by including three (or more)
outgoing eikonal lines in $W^{\rm (f)}$, Eq.\ (\ref{Wdef}).

\subsection{Resummation for the soft function}

The partonic cross section $\hat \sigma$
must be independent of the refactorization scale, $\mu$ in
         Eq.\ (\ref{refact}), at
which the soft (but still perturbative) wide-angle radiation, sensitive
only to overall color flow,  and
the truly short-distance dynamics are separated:
\be
\mu{d\over d\mu}\;
\left[\, \frac{d\hat \sigma^{\rm (f)}}{d\hat\eta}\, \right] =
0\, .
\label{muindep}
\ee
This condition applied to the right-hand side
of (\ref{refact}) leads to a renormalization group equation,
\be
\left(\mu\frac{\partial}{\partial\mu}+\beta(g_s)
\frac{\partial}{\partial g_s}\right) S^{\rm (f)}_{LI}
         = -\left({\Gamma}^{\rm (f)}_S(\hat\eta)\right)^{\dagger}_{LJ}S^{\rm
(f)}_{JI} -
S^{\rm (f)}_{LJ}\left({\Gamma}_S^{\rm (f)}(\hat\eta)\right)_{JI}\, ,
\label{rgS0}
\ee
with, as usual, $g_s=\sqrt{4\pi\alpha_s}$.
Here $\left({\Gamma}_S^{\rm (f)}(\hat\eta)\right)_{IL}$
may be thought of as an anomalous
dimension matrix \cite{KOS2,KOS1}.
These anomalous dimension matrices, and their
eigenvalues, depend on the c.m.\ rapidity, $\hat\eta$,
of the partons that participate in the hard scattering.
Equivalently, they depend upon $\hat\theta$, the scattering
angle in the partonic center-of-mass.
We always fix the observed jet at $\phi = 0$.

Because the soft function depends on the cutoff $M$ for
radiation into $\bar\O$, the $Q_\O$-dependence  is related
to the anomalous dimensions by
\be
\left(Q_\O\frac{\partial}{\partial Q_\O} - \beta(g_s)
\frac{\partial}{\partial g_s}\right) S^{\rm (f)}_{LI}
         =
   \left({\Gamma}^{\rm (f)}_S(\hat\eta)\right)^{\dagger}_{LJ}S^{\rm
(f)}_{JI} +
S^{\rm (f)}_{LJ}\left({\Gamma}_S^{\rm (f)}(\hat\eta)\right)_{JI}
- M{\partial S^{\rm (f)}_{LI}  \over \partial M}\, .
\label{chain}
\ee
To make this equation explicit, we must interpret the
derivative with respect to the cutoff.

The derivative with respect to $M$
fixes the energy flow into $\bar \O$  equal
to its maximum value, as may be seen from Eq.\ (\ref{Sdef}).
This ensures that at least one gluon with energy much
larger than $Q_\O$ appears in $\bar\O$.
Because the soft function begins at order $\alpha_s^0$,
the result may always be written as
\be
M{\partial S^{\rm (f)}_{LI} \over \partial M}
=
-\left(\Delta{\Gamma}^{\rm
(f)}_S(\hat\eta,\O)\right)^{\dagger}_{LJ}S^{\rm
(f)}_{JI} -
S^{\rm (f)}_{LJ}\left(\Delta{\Gamma}_S^{\rm 
(f)}(\hat\eta,\O)\right)_{JI}  + {\cal S}\left(\hat\eta,\frac{M}{\mu},
\frac{Q_\O}{\mu},\alpha_s(\mu)\right)\, ,
\label{parSparM}
\ee
in terms of an effective anomalous dimension matrix $\Delta \Gamma$
and a remainder term, ${\cal S}$.
Both $\Delta\Gamma$ and ${\cal S}$ are computed from real-gluon
emission, and hence
depend on the
geometry of the region $\O$ in which soft radiation
is measured.
The remainder, ${\cal S}$ begins at order $\alpha_s^2$, and in general
includes contributions proportional to $\alpha_s^2\ln Q_\O$ \cite{DasSal01}.
The logarithmic enhancements in these
contributions are associated with the ratio $M/Q_\O$.
In these contributions, energetic gluons, of energy $\sim M$ produce 
secondary radiation
that appears in $\O$.  In ${\cal S}$, this
radiation resolves the color flow of the energetic radiation from that of the
original eikonal lines that define the soft function in Eq.\ (\ref{Sdef}).
As such, we cannot absorb these contributions into an anomalous
dimension matrix for $2\rightarrow 2$ scattering, but must consider
more general color flows.  To preserve the simplicity of
a single matrix anomalous dimension, we shall simply neglect ${\cal S}$ below.
In a sense, this is equivalent to neglecting logarithms of
$M/Q_\O$ compared to those in $p_T/Q_\O$.   This is by no means a
systematic approximation, however, and we must regard the resulting
formalism as a somewhat simplified model for the actual cross section.
The study of Ref.\ \cite{DasSal02} for $\rm e^+e^-$
annihilation, however, suggests that the
effect of neglecting ${\cal S}$ does not change the behavior of the
cross section qualitatively.

These reservations notwithstanding, we now define
\be
\left({\Gamma}^{\rm
(f)}_S(\hat\eta,\O)\right)_{JI}
=
  \left({\Gamma}^{\rm (f)}_S(\hat\eta)\right)_{JI}
+
\left(\Delta{\Gamma}^{\rm
(f)}_S(\hat\eta,\O)\right)_{JI}\, ,
\ee
which depends on the
geometry of the region $\O$ in which soft radiation
is observed.
Then, neglecting ${\cal S}$, we have an equation for
the $Q_\O$ dependence of the soft matrix
\be
\left(Q_\O\frac{\partial}{\partial Q_\O} - \beta(g_s)
\frac{\partial}{\partial g_s}\right) S^{\rm (f)}_{LI}
         =
\left({\Gamma}^{\rm
(f)}_S(\hat\eta,\O)\right)^{\dagger}_{LJ}S^{\rm
(f)}_{JI}
+ S^{\rm (f)}_{LJ}\left({\Gamma}_S^{\rm (f)}(\hat\eta,\O)\right)_{JI}\, .
\label{rgS}
\ee
As discussed above,
Eq.\ (\ref{rgS}) holds at
leading logarithm in $\mu$, and hence $Q_\O/\mu$,
consistent with the accuracy of the underlying
factorizations, Eqs.\ (\ref{hatsigfact}) and (\ref{refact}),
but neglects corrections at the same power in $\ln (M/Q_\O)$. In the following, we therefore drop the argument $M$ in the cross section and soft function.

To solve Eq.\ (\ref{rgS}),
we go to a basis for the color matrices $c_I^{\rm (f)}$
that diagonalizes $\Gamma_S^{\rm (f)}(\hat\eta,\O)$,
through a transformation matrix $R$,
\be
\left(\Gamma_S^{\rm (f)}(\hat\eta,\O)
\right)_{\gamma\beta} \equiv
\lambda^{\rm (f)}_\beta(\hat\eta,\O)\delta_{\gamma\beta}=
R^{\rm (f)}_{\gamma I\, }\left(\Gamma_S^{\rm
(f)}(\hat\eta,\O)\right)_{IJ}\,
R^{\rm (f)}{}^{-1}_{J\beta},
\label{Gamdia}
\ee
where
\be
\lambda^{\rm (f)}_\beta(\hat\eta,\O)={\alpha_s\over \pi}\,
\lambda_\beta^{\rm (f,\; 1)}(\hat\eta,\O)
+\dots
\label{lambdadef}
\ee
are the eigenvalues of $\Gamma_S^{\rm (f)}(\hat\eta,\O)$.
Here and below, Greek indices $\beta, \; \gamma$ indicate that a matrix
is evaluated in the basis where
the soft anomalous dimension has been diagonalized.
Thus, for the soft and short-distance functions we also write,
\ba
S^{\rm (f)}_{\gamma\beta} &=&
\left[\left(R^{\rm (f)-1}\right)^{\dagger}\right]_{\gamma L} \;
         S^{\rm (f)}_{LK} \; \left[R^{\rm (f)-1}\right]_{K\beta}
\nonumber \\
H^{\rm (f)}_{\gamma\beta} &=& \left[R^{\rm (f)}\right]_{\gamma K} \;
H^{\rm (f)}_{KL} \; \left[R^{\rm (f)\dagger}\right]_{L\beta}.
\label{colortransform}
\ea
The transformation matrix $R^{\rm (f)-1}$ is given by the eigenvectors
of the anomalous dimension matrix,
\begin{equation}
\left( R^{(\rm f)\, -1} \right)_{K \beta}
\equiv \left( e_\beta^{(\rm f)}
\right)_K\, .
\label{Rdef}
\end{equation}

The solution to Eq.\ (\ref{rgS}), which resums leading logarithms
of $Q_\O/\mu$, is straightforward.
We introduce a combination of eigenvalues
of $\Gamma_S^{\rm (f)}$, $E_{\gamma\beta}^{\rm (f)}(\hat\eta,\O)$, given
by
\be
E^{\rm (f)}_{\gamma\beta}(\hat\eta,\O) =
\frac{2}{{\beta}_0}\left[
{\lambda}^{\rm (f,\; 1)\star}_{\gamma}(\hat\eta,\O) +
{\lambda}^{\rm (f,\; 1)}_{\beta}(\hat\eta,\O)\right]\, ,
\label{Edef}
\ee
in terms of the lowest-order coefficient
of the beta function, ${\beta}_0 = (11N_c-2n_f)/3$, for
$N_c$ colors and $n_f$ quark flavors.
The soft function is then
\ba
S^{\rm (f)}_{\gamma\beta}\left(\hat\eta,\Omega,
{Q_\O\over\mu},\alpha_s(\mu)\right)
&=&
S^{\rm
(f)}_{\gamma\beta}\left(\hat\eta,\Omega,1,\alpha_s(Q_\Omega)\right)
\ \exp \left[ E^{\rm (f)}_{\gamma\beta}(\hat\eta,\O)\; \int_\mu^{Q_\O}
{d\mu'\over
\mu'}\;
\left( {\beta_0\over 2\pi}{\alpha_s(\mu')}
\right)\right]
\nonumber\\
&=&
S^{\rm
(f)}_{\gamma\beta}\left(\hat\eta,\Omega,1,\alpha_s(Q_\Omega)\right)\
\left( {\alpha_s(\mu)\over \alpha_s(Q_\O)}\right)^{E^{\rm
(f)}_{\gamma\beta}(\hat\eta,\O)}\, ,
\label{softsoln}
\ea
where in the second form we have
reexpressed the result in terms of one-loop running couplings.
We now set the refactorization scale, $\mu$, equal to the
transverse momentum of the observed jet, $p_T$, and find,
for the partonic cross section,
\ba
\frac{{d\hat{\sigma}}^{\rm (f)}
\left(p_T,\hat\eta,\mu_F,Q_\O,\alpha_s(\mu_F)\right) }
{d\hat\eta   } &=&
\sum_{\beta,\,\gamma}\;
         H^{\rm (f)}_{\beta\gamma}\left(p_T,\hat\eta,
{\alpha}_s({\mu_F})\right)
\nonumber\\
&\ & \hspace{10mm} \times
S^{\rm
(f)}_{\gamma\beta}\left(\hat\eta,\Omega,1,\alpha_s(Q_\Omega)\right)\
\left( {\alpha_s(p_T)\over \alpha_s(Q_\O)}\right)^{E^{\rm
(f)}_{\gamma\beta}(\hat\eta,\O)}\, .
\label{HSresum}
\ea
  At leading logarithm, we approximate $H^{\rm (f)}$
by $H^{\rm (f,1)}$ and $S^{\rm (f)}$ by $S^{\rm (f,0)}$,
as observed above. Then, for $Q_\O=p_T$, this expression reduces
to the Born
cross section for inclusive jet production, as in Eq.\ (\ref{lo}).
The corresponding differential cross section in $Q_\O$,
valid to leading logarithm, is
\ba
Q_\O\frac{{d\hat{\sigma}}^{\rm (f)}\left(p_T,\hat\eta,\mu_F,
Q_\O,\alpha_s(\mu_F)\right)}
{d\hat\eta dQ_\O} &\ & \nonumber\\
&\ & \hspace{-60mm} = \sum_{\beta,\gamma}
         H^{\rm (f,\;1)}_{\beta\gamma}\left(p_T,\hat \eta,
{\alpha}_s(\mu_F)\right)\; S^{\rm (f,\; 0)}_{\gamma\beta}
\; E^{\rm (f)}_{\gamma\beta}(\hat\eta,\O)\,
\left[{\beta_0\alpha_s(Q_\O)\over 2\pi}\right]\,
\left( \frac{\alpha_s(p_T)} {\alpha_s(Q_\O)}
\right)^{E^{\rm (f)}_{\gamma\beta}(\hat\eta,\O)}\, .
\label{lla}
\ea
For consistency, we must recover at lowest order
the eikonal approximation to the complete $2\rightarrow 3$
cross section at fixed $Q_\O$ from (\ref{lla}).
We shall see how this result emerges in the
following section, where we describe the calculation
of the exponents $E_{\gamma\beta}^{\rm (f)}$.

Once the matrix $R^{\rm (f)-1}$, Eq.\ (\ref{Rdef}),
is determined in a specific
basis for color exchange, the soft and short-distance
matrices in Eq.\ (\ref{lla}) are found by applying Eq.\
(\ref{colortransform}).
The lowest-order soft matrices $ S^{\rm (f,\;0)}_{LK}$  reduce to color
traces only, while at leading logarithm,
         the hard matrices $H^{\rm (f, \;1)}_{KL}$ are given by
squares of the corresponding
leading-order partonic $2\rightarrow 2$ amplitudes,
projected onto the original color basis.

We are now ready to introduce a convenient measure of soft
radiation into region $\O$ in our dijet events.
Consider the ratio
\ba
\rho_{AB}(p_T,\eta,Q_\O,\O)
=
{1\over d\sigma_{AB}^{\rm (LO)}/d\eta d p_T  }\
{d\sigma_{AB}(p_T,\eta,Q_\O,\O)\over d\eta d p_T }\, ,
\label{rhoABdef}
\ea
where the numerator is the inclusive hadronic cross
section given in Eq.\ (\ref{inclusive}) above,
and the denominator is the lowest-order dijet
cross section for the same hadrons,
\begin{eqnarray}
\frac{d\sigma_{AB}^{\rm (LO)}}{d\eta d p_T }
      & = & \sum_{\rm f}\,
\int dx_Adx_B \;
{\phi}_{f_A/A}\left(x_A, \mu_F\right) {\phi}_{f_B/B}\left(x_B,
\mu_F\right) \nonumber \\
& \,\, & \hspace{7mm} \times \, \delta\left( p_T-{\sqrt{\hat s}\over
2\cosh\hat\eta}\right)\,
\sum_{I,\,L}\;
         H^{\rm (f,\; 1)}_{IL}\left(p_T,\hat\eta,
{\alpha}_s(\mu_F)\right)
S^{\rm (f,\; 0)}_{LI}\, .
\label{LOdijet}
\end{eqnarray}
Here we have used Eq.\ (\ref{lo}) to relate the LO
hard-scattering function $\hat \sigma$, in collinear
factorization,
to the lowest-order, refactorized hard and soft functions,
$H$ and $S$.
By comparing this expression with
Eqs.\ (\ref{inclusive}) and (\ref{HSresum})
for the numerator in (\ref{rhoABdef}), we see that,
so long as $\mbox{ Re } E_{\gamma \beta} > 0$,
$\rho_{AB}$ vanishes for small $Q_\O$ (strictly
speaking, at $Q_\O=\Lambda_{\rm QCD}$), increases
monotonically with $Q_\O$, and reaches unity at $Q_\O=p_T$.
This behavior, due to resummation, is to be contrasted
with the corresponding NLO expression for the same
quantity, which diverges toward minus infinity
for vanishing $Q_\O$, where the virtual contribution
dominates.  The quantity $\rho_{AB}$ predicts the distribution of
bremsstrahlung
into region $\O$.

\section{The Anomalous Dimension Matrices}

We now turn to the calculation of
the anomalous dimensions $\Gamma_S^{\rm (f)}$,
introduced in Eq.\ (\ref{rgS}) above.
Equation (\ref{rgS}), in turn, followed directly from the
refactorization of the hard-scattering function,
Eq.\ (\ref{refact}), with an extra contribution
from real-gluon emission into $\bar \O$, as in
Eqs.\ (\ref{chain}) and (\ref{parSparM}).
The refactorization scale, $\mu$
of Eq.\ (\ref{refact}) was identified with the
renormalization scale for the operators $W$ in the definition
of the soft functions $S^{\rm (f)}$, Eq.\ (\ref{Sdef}).
In this section, we shall derive these anomalous dimensions
from the renormalization of the soft function.
To compute
the anomalous dimensions at one loop, we may identify
the cutoff $M$ for real gluon
emission  in the definition of $S^{\rm (f)}$
with the refactorization scale, $\mu$.  We thus suppress
the additional argument $M$ in $S$ for this calculation.
We will work in Feynman gauge.

The renormalization of multi-eikonal
vertices has been discussed in  some detail in Ref.\
\cite{KOS2},
and we shall follow the method
outlined there, with an important difference.
In \cite{KOS2}, the anomalous dimensions
$\Gamma_S^{\rm (f)}$ were computed in axial gauge,
after dividing an eikonal scattering amplitude by
eikonal self-energy functions.  This extra factorization
eliminated double poles in dimensional
regularization, which are associated with collinear
emission by the nonabelian phase operators.  In
axial gauge, these singularities appear only
in self-energy diagrams.
In the present case, such an extra factorization
is not necessary, because
         $S_{LI}^{\rm (f)}$, defined as in (\ref{Sdef}) is
free of  collinear singularities
associated with the eikonal lines, after the
sum over intermediate states.
The remaining ultraviolet divergences
in $S_{LI}^{\rm (f)}$ may be compensated
through local counterterms, just as in \cite{KOS2}.
       From these counterterms,
we may simply read off the
entries of the anomalous dimension matrix
$\Gamma^{\rm (f)}_S$ in an MS
renormalization scheme.

Given the definition of the
soft function, Eq.\ (\ref{Sdef}),
the counterterms for $S$
in Eq.\ (\ref{refact})
are calculated only after a sum over
final states.  They therefore
depend on the choice of $\Omega$, and the
scattering angle of the underlying process.
The following two subsections describe the one-loop calculation
of $S^{\rm (f)}_{LI}$ in this approach.

\subsection{Renormalization and color mixing}

\begin{figure}
%\vbox{\vskip 1 true in}
\centerline{\epsfxsize=9cm \epsffile{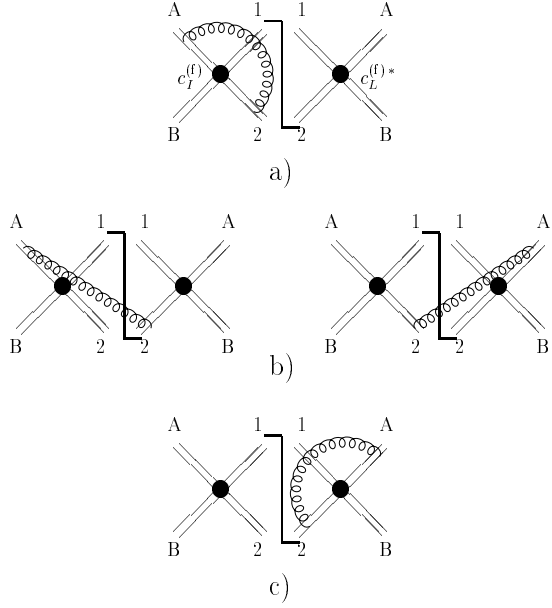}}
\caption{Diagrams for the calculation of the
anomalous dimension matrix, as in Eq.\ ({\protect \ref{omegaVint}}).
The double lines represent eikonal propagators, linked
by vertices $c_I^{\rm (f)}$ and $c_L^{\rm (f)}{}^*$,
in the amplitude and its complex conjugate.  The
vertical line represents the final state.}
\label{examplefig}
\end{figure}

To begin, we consider the diagrams
shown in Fig.\ \ref{examplefig}.
Figs.\ \ref{examplefig}a and \ref{examplefig}c
are virtual corrections,
while \ref{examplefig}b shows the corresponding
diagrams for real gluon
emission.  To compute the counterterms
associated with these diagrams, we fix the transverse energy flow
$Q_\O=0$
in Eq.\ (\ref{Sdef}).

         As is characteristic of eikonal
diagrams, Figs.\ \ref{examplefig}a-c
are all proportional to scaleless integrals,
which vanish in dimensional regularization.
At $Q_\O=0$, therefore, $S^{\rm (f,1)}_{LI}$ is
given entirely by its one-loop counterterms,
which we denote as \cite{KOS1}
\ba
S_{LJ}^{(\rm f,1)}\left(\hat\eta,\Omega,0,\alpha_s(\mu),\varepsilon\right)
&=&
- \left(Z_S^{({\rm f},\,
1)}\left(\hat\eta,\Omega,\alpha_s(\mu),\varepsilon\right)
\right)^\dagger_{LI}\,
S_{IJ}^{(\rm f,0)}
\nonumber\\
&\ & \quad
-
S^{(\rm f,0)}_{LI}\,
\left(Z_S^{({\rm f},\, 1)}
\left(\hat\eta,\Omega,\alpha_s(\mu),\varepsilon\right)\right)_{IJ}\,
,
\label{Sren}
\ea
where $S^{(0)}$ is the zeroth order matrix, which,
by Eq.\ (\ref{Sdef}) is a set of color factors,
independent of $\hat\eta$ and $\O$. The elements of $Z$ are computed in
$D=4-\varepsilon$ dimensions.   It is important to note that
the $\varepsilon$ poles, given by the one-loop counterterms,
have the interpretation of {\it infrared}, rather than
ultraviolet, divergences in $S_{LI}^{\rm (f)}$,
and that they cancel infrared divergences from real-gluon emission
into region $\O$.\footnote{More specifically, the divergences
from virtual diagrams are proportional to $\delta(Q'_\O)$
in {\protect (\ref{Sdef})}, and cancel real-gluon
emission as integrable distributions.}
Following the standard analysis, the one-loop
anomalous dimensions are given by
\begin{equation}
\left(\Gamma_S^{\rm (f)}(\hat\eta,\Omega)\right)_{IJ}  =
- \alpha_s
\frac{\partial}{\partial \alpha_s}
\mbox{ Res}_{\varepsilon \rightarrow 0}
\left(
Z_S^{\rm (f,1)}\left(\hat\eta,\Omega,\alpha_s(\mu),\varepsilon\right)
\right)_{IJ}
\label{residue}
\, .
\end{equation}
The calculation of the $Z$'s is thus essentially
equivalent to the calculation of the anomalous dimensions.
To carry out these calculations, however, we must specify
a basis of color tensors, $c_I$.

For the two-to-two partonic processes, Eq.\ (\ref{pprocess}),
we denote by $r_i$ the color index associated with
parton (or eikonal line) $i=A,B,1,2$.
As above,  capital letters, $I,J\dots$ are indices in the
space of color exchanges,
spanned by a chosen basis for each partonic process.
We will restrict the discussion here to the
calculation of anomalous dimensions when the
incoming partons are quarks or antiquarks.
Convenient bases for describing the color flow for quark
processes are $t$-channel singlet-octet
bases, given by
\begin{eqnarray}
c_1 & = & \delta_{r_A,\,r_1} \delta_{r_B,\,r_2}, \nonumber \\
c_2 & = & - \frac{1}{2 N_c} \delta_{r_A,\,r_1} \delta_{r_B,\,r_2} +
\frac{1}{2} \delta_{r_A,\,r_B} \delta_{r_1,\,r_2}
\label{qqbarbas}
\end{eqnarray}
for $q \bar{q} \rightarrow q \bar{q}$, and
\begin{eqnarray}
c_1 & = & \delta_{r_A,\,r_1} \delta_{r_B,\,r_2}, \nonumber  \\
c_2 & = & - \frac{1}{2 N_c} \delta_{r_A,\,r_1} \delta_{r_B,\,r_2} +
\frac{1}{2} \delta_{r_A,\,r_2} \delta_{r_B,\,r_1}.
\label{qqbas}
\end{eqnarray}
for $q q \rightarrow q q$. The process $q \bar{q} \rightarrow g g$ is best
described in terms of the color basis
\begin{eqnarray}
c_1 & = & \delta_{r_1,\,r_2}\, \delta_{r_A,\,r_B} , \nonumber \\
c_2 & = & d_{r_2r_1c} \left(T^c_F\right)_{r_Br_A},
\label{ggqbas} \\
c_3 & = & i f_{r_2r_1c} \left(T^c_F\right)_{r_Br_A}, \nonumber
\end{eqnarray}
where $c_1$ is the $s$-channel singlet tensor, and $c_2$ and $c_3$ are the
symmetric and antisymmetric octet tensors, respectively.
Bases for other possible partonic processes with gluon eikonal lines
have been given, for example, in Refs.\ \cite{Oderda,KOS2}.

Let $Z^{(ij)}$ denote the contribution to the
counterterms from all the one-loop graphs in which
the gluon connects eikonal lines $i$ and $j$.  In
this notation, the calculation of Fig.\ \ref{examplefig}
gives us $Z^{(A2)}$.  We emphasize that the $Z$'s,
and hence the anomalous dimensions, are to be calculated
only after combining real-gluon emission diagrams
with virtual diagrams.  The $Z$'s are
constructed to give local counterterms that cancel
only those ultraviolet divergences that are left over
after the real-virtual cancellation has been carried out.

\begin{figure}
%\vbox{\vskip 1 true in}
\centerline{\epsfysize=3cm \epsfbox[71 356 545 478]{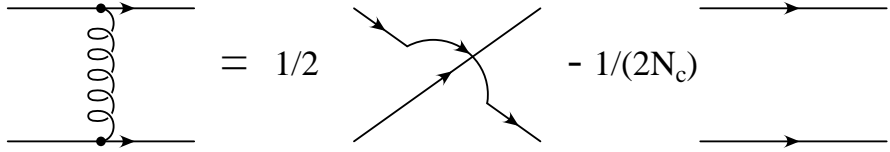}}
\caption{Color identity corresponding to Eq.\ ({\protect \ref{identity})}.}
\label{colorfig}
\end{figure}

To find the $Z^{(ij)}_{IJ}$'s in
the color bases (\ref{qqbarbas}) and (\ref{qqbas}),
we rewrite the various one-loop virtual diagrams
in terms of the original color basis, using the identity
shown in Fig. \ref{colorfig},
\begin{equation}
T^a_{ij} T^a_{kl} = \frac{1}{2} \left( \delta_{il} \delta_{jk} -
\frac{1}{N_c} \delta_{ij} \delta_{kl} \right)\,
\label{identity}
\end{equation}
for quark processes.  For scattering processes involving gluons,
many useful identities can be found, for example, in \cite{mac}.
This procedure results in a $2\times 2$ matrix decomposition
for scattering
involving only quark
and antiquark eikonal lines,
describing the mixing under renormalization of their
color exchanges.
The annihilation of a pair of quark and antiquark
eikonals into two gluon eikonal
lines gives a $3\times 3$ matrix structure, while for incoming and
outgoing gluonic eikonals,
we get an $8\times 8$ matrix \cite{KOS2}.

For a given $Z^{(ij)}$,
the momentum-space integral appears as an overall factor.
It is then convenient to introduce the notation
\ba
\left(Z_S^{(ij)}\right)_{LI}
=
\zeta^{(ij)}_{LI}\; \omega^{(ij)}\, ,
\label{factorZ}
\ea
where the $\zeta_{LI}$ are the coefficients that
result from the color decomposition of the virtual
diagram, and where the $\omega^{(ij)}$s include the ultraviolet
  pole part of the
momentum space integral for the soft function $S$ and
remaining overall constants.
Defined in this fashion, the sign of each $\omega^{(ij)}$
depends on the flow of flavor in the underlying
process (see below). From (\ref{residue}) and (\ref{factorZ})
the relation between the $\omega$'s and the anomalous dimension matrices
is
\begin{equation}
\left( \Gamma_S^{(\rm f)} \right)_{LI} = - \varepsilon 
\sum_{\{(ij)\}} \zeta^{(ij)}_{LI}
  \;\omega^{(ij)}\;. \label{fulladm}
\end{equation}

As an example, we give the color decomposition of
$\Gamma^{(A2)}$ in $q \bar{q}
\rightarrow q \bar{q}$, for which we find, using Eq.\ (\ref{identity}),
\begin{eqnarray}
\zeta^{(A2)}  = \left( \begin{array}{cc} 0 &
\frac{C_F}{2 \,N_c}  \\
1 & - \frac{1}{N_c}  \end{array} \right)\, .
\end{eqnarray}
The step to the anomalous dimensions is trivial, following Eqs.\
(\ref{residue}) and (\ref{fulladm}).
We define
\be
\Gamma^{(ij)} \equiv -\varepsilon\, \omega^{(ij)}\, .
\label{Gamom}
\ee
Then we may  summarize the full anomalous dimension matrices
(\ref{fulladm})
in terms of the following combinations,
\begin{eqnarray}
\alpha^{\rm (f)} & \equiv & \Gamma^{(AB,\, \rm f)} +
\Gamma^{(12,\, \rm f)}, \nonumber \\
\beta^{\rm (f)} & \equiv & \Gamma^{(A1,\, \rm f)} +
\Gamma^{(B2,\, \rm f)}, \nonumber \\
\gamma^{\rm (f)} & \equiv & \Gamma^{(A2,\, \rm f)} +
\Gamma^{(B1,\, \rm f)}.
\label{abgdef}
\end{eqnarray}
In the basis (\ref{qqbarbas}), the anomalous dimension matrix for the
process $q \bar{q} \rightarrow q \bar{q}$ is given by
\begin{equation}
\Gamma^{(q\bar q \rightarrow q \bar{q})}_S = \left( \begin{array}{cc}
C_F \beta^{(q \bar q \rightarrow q \bar q)} &
\frac{C_F}{2 N_c} \left(
\alpha^{(q \bar q \rightarrow q \bar q)} +
   \gamma^{(q \bar q \rightarrow q \bar q)} \right) \\
\alpha^{(q \bar q \rightarrow q \bar q)} +
   \gamma^{(q \bar q \rightarrow q \bar q)} \quad &
C_F \alpha^{(q \bar q \rightarrow q \bar q)} - \frac{1}{2 N_c}
\left( \alpha^{(q \bar q \rightarrow q \bar q)} +
\beta^{(q \bar q \rightarrow q \bar q)}
+ 2 \gamma^{(q \bar q \rightarrow q \bar q)} \right) \end{array} \right)\, .
\label{qqbarG}
\end{equation}
Here and below,
we have suppressed the arguments of
the momentum factors, $\alpha,\beta$ and $\gamma$ for compactness, but they all
depend on $\hat{\eta}$ and on $\O$.
The color decomposition for the process $q \bar q \rightarrow \bar q q$ is
the same as above, Eq. (\ref{qqbarG}). The momentum
parts change, however:
$\beta^{(q \bar q \rightarrow \bar q q)} = -
\gamma^{(q \bar q \rightarrow q \bar q)}$ and
$\gamma^{(q \bar q \rightarrow \bar q q)} =
- \beta^{(q \bar q \rightarrow q \bar q)}$.

Similarly, for $q q \rightarrow q q$
in the basis (\ref{qqbas}), we have
\begin{equation}
\Gamma_S^{(q q \rightarrow q q)} =
\left( \begin{array}{cc} C_F \beta^{(q q \rightarrow q q)} &
\frac{C_F}{2 N_c} \left(
\alpha^{(q q \rightarrow q q)} +
\gamma^{(q q \rightarrow q q)} \right) \\
\alpha^{(q q \rightarrow q q)} +
\gamma^{(q q \rightarrow q q)}  \quad &  C_F \gamma^{(q q \rightarrow q q)} -
\frac{1}{2 N_c}
\left( 2 \alpha^{(q q \rightarrow q q)} +
\beta^{(q q \rightarrow q q)} + \gamma^{(q q \rightarrow q q)} \right)
\end{array} \right)\, .
\label{qqG}
\end{equation}
Note that certain momentum factors for quark-quark and quark-antiquark differ
by signs, due to the eikonal Feynman rules (see below, Sec.\ 3.2).

For $q \bar{q} \rightarrow g g$ we find in the basis (\ref{ggqbas}),
\begin{equation}
\Gamma_S^{(q \bar{q} \rightarrow gg)} = \left( \begin{array}{ccc}
C_F \Gamma^{(AB)} + C_A \Gamma^{(12)} \,\, & 0 & \frac{1}{2} \left(
\beta +\gamma \right) \\
       0 & \xi \,\, &
       \frac{N_c}{4} \left( \beta + \gamma \right) \\
\beta + \gamma   &  \frac{N_c^2 - 4}{4 N_c} \left( \beta + \gamma \right)
       & \xi \end{array} \right),  \label{qqggADM}
\end{equation}
where
$\xi  \equiv  \frac{N_c}{4} \left( \beta - \gamma \right) - 
\frac{1}{2 N_c} \Gamma^{(AB)} + \frac{N_c}{2}  \Gamma^{(12)}$.
Here we have suppressed the superscripts, and define  the
$\Gamma^{(ij)}$, as well as $\alpha$, $\beta$ and $\gamma$
for $q \bar{q} \rightarrow g g$
to be the same as those
above for $q \bar{q} \rightarrow
q \bar{q}$.  We have done so because, for
processes involving gluons, the factorization
into momentum and color factors
as in (\ref{factorZ}) is ambiguous.  The gluon-eikonal gluon
vertex has both momentum and color parts,
and the sign of each depends on the orientation with
which the exchanged gluon attaches to the gluon eikonal line.
Their product, and the anomalous dimension matrix, of course,
is independent of the way the
graphs are drawn.

\subsection{Momentum integrals} \label{momentum}

We now compute the momentum integrals
$\omega^{(ij)}$ that contribute to the
matrix of renormalization constants
$\left(Z_S^{\rm (f)}\right)_{LI}$,
at one loop, through Eq.\ (\ref{factorZ}).
The momentum factor for the virtual diagram in
which the gluon line connects eikonal lines $i$ and $j$,
such as Fig.\ \ref{examplefig}a,
may be written as  \cite{KOS2,KOS1}
\ba
\omega^{(ij)}_V
\left(\delta_i\beta_i,\delta_j\beta_j,\Delta_i,\Delta_j\right)
&=&
g_s^2\, \int\limits_{P.P.} {d^nk \over (2\pi)^n}\,
{-i\over k^2+i\epsilon}\;
{\Delta_i\Delta_j\beta_i\cdot\beta_j
\over
\left(\delta_i\beta_i\cdot k+i\epsilon\right)
\left(\delta_j\beta_j\cdot k+i\epsilon\right)}\, ,
\label{omegaVint}
\ea
where here and below the subscript $P.P.$ indicates that the integral is
defined by its ultraviolet pole part.
In (\ref{omegaVint}), $\delta_i=1(-1)$ for momentum
$k$ flowing in the same (opposite) direction as the momentum
flow of line $i$, and  $\Delta_i=1(-1)$ for $i$ a quark (antiquark)
line,
while, as noted above, we define the
momentum integrals for gluons to have the same signs as for quark-antiquark
scattering.
      For example,
Fig.\ \ref{examplefig}a for $q \bar{q} \rightarrow q \bar{q}$ has
$i$ = A, $j$ = 2, with $\Delta_A=1$,
$\Delta_2=-1$ and with $\delta_A=\delta_2$.
In this notation, each of the real gluon emission diagrams,
such as in Fig.\ \ref{examplefig}b, takes the form
\ba
\omega^{(ij)}_R
\left(\delta_i\beta_i,\delta_j\beta_j,\Delta_i,\Delta_j\right)
&=&
\, g_s^2\, \int\limits_{P.P.} {d^nk \over (2\pi)^{n-1}}\,
\delta_+(k^2)\; \left(1- \Theta(\vec k)\right)\;
{\Delta_i\Delta_j\beta_i\cdot\beta_j
\over
\left(\delta_i\beta_i\cdot k\right)
\left(\delta_j\beta_j\cdot k\right)}\, ,
\label{omegaRint}
\ea
with $\delta_i$ and $\delta_j$ defined by the corresponding
virtual diagram. The function $\Theta(\vec k) = 1$ when
the vector $\vec k$
is directed into region $\O$,
and is zero otherwise.  Both real-emission diagrams give the
same answer, while the two virtual diagrams are complex
conjugates of each other.

The sum of the virtual diagram in the amplitude with
either of the real diagrams gives:
\ba
\omega^{(ij)}\left(\delta_i\beta_i,\delta_j\beta_j,\Delta_i,\Delta_j\right)
&=& \omega^{(ij)}_V + \omega^{(ij)}_R
\nonumber\\
&=& - g_s^2\, \int\limits_{P.P.} {d^nk \over (2\pi)^{n-1}}\,
\delta_+(k^2)\; \Theta(\vec k)\;
{\Delta_i\Delta_j\beta_i\cdot\beta_j
\over
\left(\delta_i\beta_i\cdot k\right)
\left(\delta_j\beta_j\cdot k\right)}
\nonumber\\
&\ & \quad
+ \Delta_i\Delta_j\delta_i\delta_j\; {\alpha_s\over
2\pi}\,
{i\pi\over\varepsilon}\, \left(1-\delta_i\delta_j\right)
\nonumber\\
&\equiv& - {\Gamma^{(ij)}\over \varepsilon}\, .
\label{W1loop}
\ea
In the last line, we have recalled the definition
(\ref{Gamom}).
The term with the $k$-integral is the real
remainder of the cancellation of the real and
virtual diagrams, while the imaginary part is
nonzero only for $(ij)=(12)$ or $(AB)$.  At $Q_\O=0$,
the real diagrams $\omega_R^{(ij)}$ get contributions
only from gluon emission outside of $\O$,
and cancel contributions from $\omega^{(ij)}_V$
as shown in Eq.\ (\ref{W1loop}).  The counterterms
are identified after this cancellation has
been carried out, and are implemented as
local subtractions at the color vertices, $c_I$
and $c_L^*$.  This requires that the counterterms,
and therefore the anomalous dimensions,
be complex.

Given the anomalous dimensions computed in this fashion,
and the lowest-order hard and soft functions, we
find the differential cross section for  interjet
radiation into region $\O$ from Eq.\ (\ref{lla}).
Note that for $Q_\O\ne 0$, the cross  section
may be fixed uniquely at one loop  by the condition that the integral
of the real part down to $Q_\O=0$ must cancel the divergences of
the virtual diagrams,
that is, the counterterms identified above.
We recognize that the result is consistent with
    Eq.\ (\ref{lla}) expanded to one loop,
after a transformation to the
diagonal basis for color exchange.
The inclusive cross sections are given similarly,
by Eq.\ (\ref{HSresum}).

\section{Out-of-plane Radiation}

As an illustration of our method, we introduce a region $\Omega$ that
is fixed out of the scattering plane, as a ``rectangular
patch"
in the space of rapidity $y$ and azimuthal  angle
$\phi$.  Other choices are possible of course, and
will be explored elsewhere.

Taking the observed jet at $\phi=0$, and
the recoil jet at $\phi=\pi$, we define $\Omega$ in the hadronic center-of-mass
frame by
\ba
{\phi}_{min} < & \phi & < \; {\phi}_{max} \,  ,
\nonumber \\
      {\eta}_{min} < & y & <  {\eta}_{max}\,,
\label{patchdef}
\ea
where the boundaries are chosen such that the patch is located
outside the incoming beams and outgoing jets.
The distribution of energy radiated into the out-of-plane patch is
determined by
the eigenvectors and eigenvalues of the anomalous dimension matrices,
   $\Gamma_S^{\rm (f)}$,
Eqs.\ (\ref{qqbarG}), (\ref{qqG}), and (\ref{qqggADM}) above.
Given the definition (\ref{patchdef})
for $\O$,  and the expressions (\ref{W1loop}) for the diagrammatic
contributions, we simply need to evaluate the one loop integrals over the
loop momenta
restricted to this phase space.

\subsection{Entries for the anomalous dimensions}

For the patch region of Eq.\ (\ref{patchdef}),
all the phase space integrals of Eq.\ (\ref{W1loop}),
and hence all the entries of the anomalous dimensions, $\Gamma_S^{\rm
(f)}$,
in Eqs.\ (\ref{qqbarG})-(\ref{qqggADM})
      can be expressed in terms
of the following combinations of logarithmic and dilogarithmic functions,
\be
I(x,y,z) = \arctan(z)\, \ln\left(\frac{x+z}{y+z}\right)
+ \frac{1}{2i} \left\{ {\mbox Li}_2 \left( \frac{x+z}{z+i} \right) +
         {\mbox Li}_2 \left(\frac{y+z}{z-i}\right) \right\} -
\frac{1}{2i} \left
\{ x \leftrightarrow y \right \}\, . \label{Idef1}
\ee
We note that this function has a finite limit to $z=\pm i$, where
singularities of the dilogarithms cancel those of the arctangent:
\be
I(x,y,\pm i)={1\over \pm 2i}\left\{\, {1\over 2} \left[ \ln^2\left(
\frac{x \pm i}{\pm 2 i} \right) - \ln^2\left(\frac{y \pm i}{\pm 2 i}
\right)\right]
+ {\mbox Li}_2\left(
\frac{x \pm i}{\pm 2 i} \right) -
{\mbox Li}_2\left(\frac{y \pm i}{\pm 2 i} \right) \, \right\}\, .
\label{Idef2}
\ee
For the anomalous dimension matrices relevant in the valence quark
approximation,
we need explicit expressions for $\Gamma^{(AB,\, \rm f)}$
and $\Gamma^{(12,\, \rm f)}$,
in addition to
$\alpha^{\rm (f)}$,
$\beta^{\rm (f)}$ and
$\gamma^{\rm (f)}$, defined in Eq.\ (\ref{abgdef}).
With variables in (\ref{patchdef}) transformed to the
  partonic center-of-mass frame, the relevant arguments of the $I$'s in Eq.
(\ref{Idef1}-\ref{Idef2}) are:
\ba
x_{min} &=& \tanh\left(\frac{\eh + \hat{\eta}_{min}}{2}\right) \,
\tan(\fma/2),
\nonumber
\\
x_{max} &=& \tanh\left(\frac{\eh + \hat{\eta}_{max}}{2}\right) \,
\tan(\fma/2),
\nonumber
\\
y_{min} &=& \coth\left(\frac{\eh - \hat{\eta}_{min}}{2}\right) \,
\tan(\fma/2),
\nonumber
\\
y_{max} &=& \coth\left(\frac{\eh - \hat{\eta}_{max}}{2}\right) \,
\tan(\fma/2),
\nonumber
\\
z &=& - \frac{\sinh(\eh) + i}{\cosh(\eh)} \, \tan(\fma/2).
\ea
In these terms, we have
\ba
\frac{2{\pi}^2}{{\alpha}_s} \; \Gamma^{ (AB,\, q \bar q \rightarrow
q \bar q)}
& = &  (\hat{\eta}_{max}-\hat{\eta}_{min})(\fma-\fmi) - 2 {\pi}^2i,
\nonumber \\
\frac{2{\pi}^2}{{\alpha}_s} \; \Gamma^{(12,\, q \bar q \rightarrow
q \bar q)} & = & \{ \;
I(x_{min},x_{max},0)-I(x_{min},x_{max},z)-I(x_{min},x_{max},z^*)
         \; \} - \{ \fma \rightarrow \fmi \} \nonumber \\
& + & \{ \; -
I(y_{min},y_{max},0)+I(y_{min},y_{max},z)+I(y_{min},y_{max},z^*)
         \; \} - \{ \fma \rightarrow \fmi \} \nonumber \\
& - & 2{\pi}^2i, \nonumber \\
\frac{2{\pi}^2}{{\alpha}_s} \;
\beta^{ (q\bar q \rightarrow q \bar q)} & = &
\{ I(-y_{min},-y_{max},0)- 2 \, I(-y_{min},-y_{max},-\tan(\fma/2)) \}
- \{ \fma \rightarrow \fmi \}
\nonumber \\
&+& \{ I(x_{min},x_{max},0)- 2 \, I(x_{min},x_{max},\tan(\fma/2)) \}
- \{ \fma \rightarrow \fmi \}\, ,
\ea
while $\gamma^{(q\bar q \rightarrow q \bar q)}$
is found from $\beta^{ (q\bar q \rightarrow q \bar q)}$
through the transformation,
\be
\gamma^{ (q\bar q \rightarrow q \bar q)}(\hat{\eta}_{min},
\hat{\eta}_{max},\fmi,\fma,\eh)
  \; = \; \beta^{
(q\bar q \rightarrow q \bar q)}(-\hat{\eta}_{min},-
\hat{\eta}_{max},\fmi,\fma,-\eh)\, .
\ee
Expressions for quark-quark scattering differ only by overall signs,
determined by the factors $\Delta_i$ in Eq.\ (\ref{W1loop}).

\begin{figure}
%\vbox{\vskip 1 true in}
\centerline{\epsfysize=6.2cm \epsffile{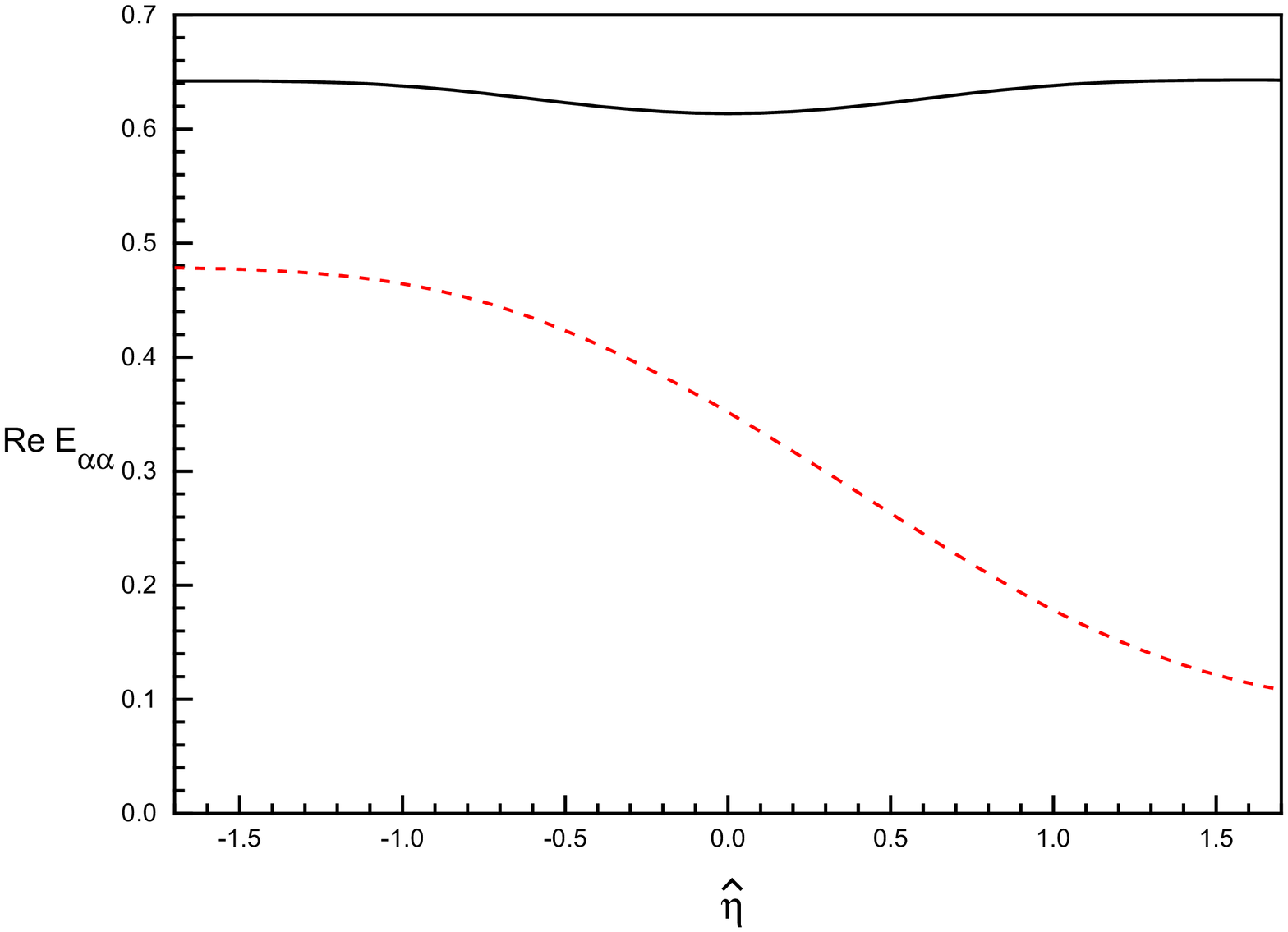} \epsfysize=6.2cm
\epsffile{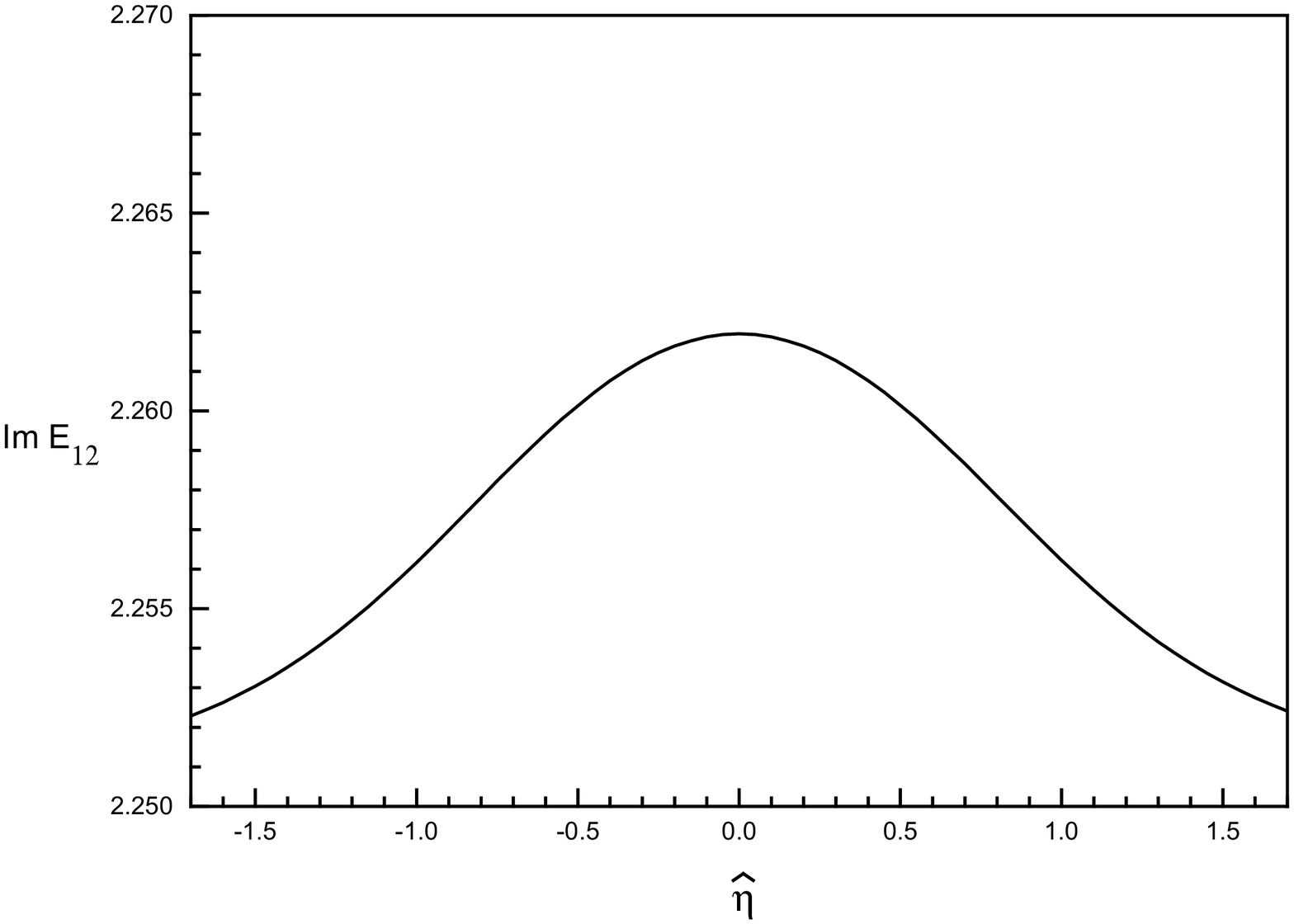} }
\caption{Real and imaginary parts of $E_{\alpha\beta}$ for
$q \bar{q} \rightarrow q \bar{q}$ with $\O$ chosen as in Eq.
({\protect \ref{patchdef}}), with $\phi_{min} = \pi/4$, $\phi_{max}=3\pi/4$,
and $\eta_{max} = -\eta_{min} = 1$.
For the real parts we plot only the diagonal elements,
${\protect \mbox{Re }} E_{\alpha \alpha}$, since
the interference terms, ${\protect \mbox{Re }} E_{\alpha \beta}$
with $\alpha\ne\beta$,
are just the averages of the former.  Note
${\protect \mbox{Im }} E_{\alpha \alpha}=0$, and that for an
$n\times n$ anomalous dimension matrix, only $n-1$
imaginary parts of the $E_{\alpha\beta}$ with $\alpha\ne \beta$
are independent.
}
\label{Eabfig}
\end{figure}

\begin{figure}
%\vbox{\vskip 1 true in}
\centerline{\epsfysize=6.2cm \epsffile{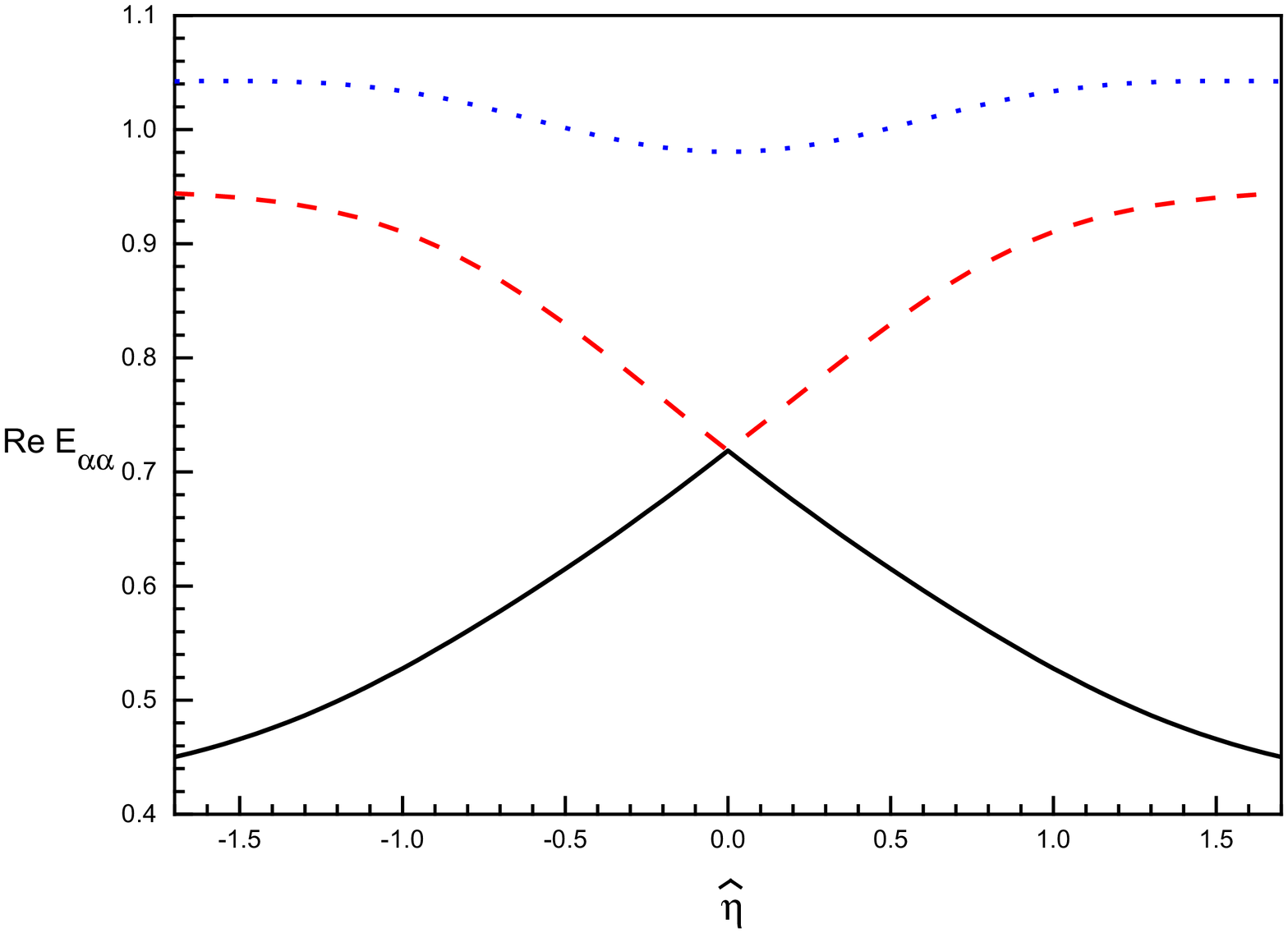} \epsfysize=6.2cm
\epsffile{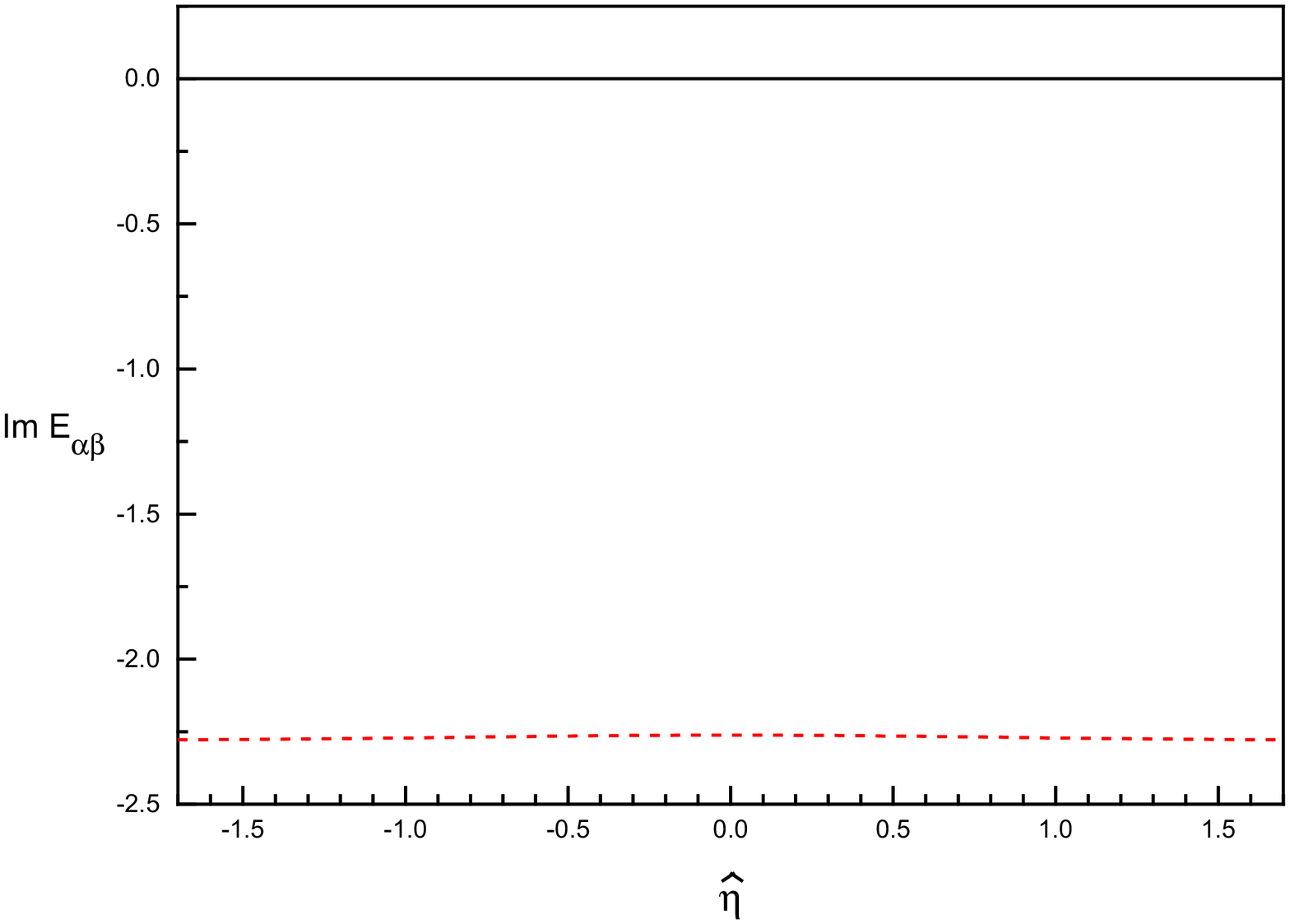} }
\caption{Real and imaginary parts of $E_{\alpha\beta}$ for
$q \bar{q} \rightarrow g g$.
}
\label{Eabfigqqgg}
\end{figure}

The anomalous dimension matrices are found by
substituting these results into Eq.\ (\ref{qqbarG}) for $q\bar q$
scattering into $q \bar q$,
or (\ref{qqggADM}) for their annihilation into
gluons,
      or (\ref{qqG})  for the quark-quark case.  It is then
straightforward to find the corresponding eigenvalues,
$\lambda^{\rm (f)}_\alpha$, Eq.\ (\ref{lambdadef}),
and transformation matrices $R^{\rm (f)-1}$,
Eq.\ (\ref{Rdef}).  The results are somewhat cumbersome, and
we do not present them here.  In Figs.\ \ref{Eabfig} and
\ref{Eabfigqqgg}, we
plot the real and imaginary parts of the exponents
$E_{\alpha\beta}$, Eq.\ (\ref{Edef}), for $q \bar q \rightarrow
q \bar q$ and $q \bar q \rightarrow  g g$,
  as functions of $\hat\eta$, the rapidity
of the observed jet in the partonic center-of-mass, for
a typical out-of-plane patch $\O$, Eq.\ (\ref{patchdef}).
For these plots we
have chosen a configuration with
$x_A = x_B$ in the factorization, Eq. (\ref{inclusive}),
so that the partonic and hadronic c.m. frames coincide, with
$\eta_{max}=-\eta_{min}=1$, $\phi_{min}=\pi/4$
and $\phi_{max}=3\pi/4$.  Note that $E_{\alpha \beta}$ changes
with $x_A$ and $x_B$ as $\O$, fixed in the hadronic c.m.,
is boosted in the partonic c.m.
The real parts of $E_{\alpha\beta}$ are all
of order unity in the  range of $\hat\eta$ shown.
We expect them to be stable against higher-order corrections,
so long as the size of the patch is of order unity.
For very small patches, we would in general anticipate
large logarithms associated with an incomplete cancellation
between real and virtual corrections.

The only remaining ingredients
necessary to compute the resummed cross sections, Eqs.\
(\ref{HSresum}) and (\ref{lla})
are the hard and soft functions at lowest order, to which we now turn.

\subsection {Hard and soft matrices at LO}

Explicit hard and soft matrices in color
space at LO for the partonic processes $q{\bar q} \rightarrow q{\bar q}$ and
$qq\rightarrow qq$
have been given in Ref.\ \cite{Oderda}.
We exhibit them here, along with those for $q{\bar q} \rightarrow g g$,
with a trivial change in overall normalization relative to \cite{Oderda}.
To use them in the resummed cross sections, Eqs.\ (\ref{HSresum})
or (\ref{lla}), it is necessary to
apply the transformation rules (\ref{colortransform}),
to take the results quoted below to the $\O$-dependent diagonal basis.

\subsubsection{Hard and soft matrices for $q{\bar q}\rightarrow q{\bar
q}$ at LO}

In the basis (\ref{qqbarbas}) the decomposition of
the Born level hard scattering for
same-flavor $q \bar{q} \rightarrow q \bar{q}$
in color space is given by the matrix
\begin{equation}
H^{(1)}\left(\hat{t},\hat{s},\alpha_s \right)=\frac{1}{N_c^2}
\,
\frac{\pi \alpha^2_s}{\hat s} \, \frac{4\hat{t}\hat{u}}{\hat{s}^2} \left(
\begin{array}{cc}
\left(\frac{C_F}{N_c}\right)^2 \chi_1  & \frac{C_F}{N_c^2} \chi_2
\vspace{2mm} \\
\frac{C_F}{N_c^2} \chi_2 & \chi_3
\end{array}
\right) \,, \label{hardqqb}
\end{equation}
where $\chi_1$,
$\chi_2$, and $\chi_3$  are
defined by
\begin{eqnarray}
\chi_1&=&\frac{\hat{t}^2+\hat{u}^2}{\hat{s}^2}      \nonumber \\
\chi_2&=&N_c \frac{\hat{u}^2}{\hat{s}\hat{t}}-
\frac{\hat{t}^2+\hat{u}^2}{\hat{s}^2} \nonumber \\
\chi_3&=&\frac{\hat{s}^2+\hat{u}^2}{\hat{t}^2}+
\frac{1}{N_c^2}\frac{\hat{t}^2+\hat{u}^2}{\hat{s}^2}
-\frac{2}{N_c}\frac{\hat{u}^2}{\hat{s}\hat{t}}\,.
\end{eqnarray}
The step to the unequal-flavor cases,
$q\bar q'\rightarrow q\bar q'$ and $q\bar q\rightarrow q'\bar q'$
     is straightforward, by dropping the $s$-channel terms for
the former, and the $t$-channel contributions for the latter.
The matrix for $q \bar q \rightarrow \bar q q$ is found from
(\ref{hardqqb}) by exchanging $\hat{t}$ and $\hat{u}$.
The lowest-order soft matrix is just given by the trace of the color
basis, $S^{(0)}_{LI}=\mbox{ Tr }(c_L^{\dagger}c_I)$, resulting in
\begin{equation}
S^{(0)}=\left(
\begin{array}{cc}
N_c^2  & 0  \\
0 &  \frac{1}{4}(N_c^2-1)
\end{array}
\right) \, . \label{softqqb}
\end{equation}

\subsubsection{Hard and soft matrices for $q q \rightarrow q q$ at LO}

>For this process we get, in the basis Eq. (\ref{qqbas}),
a hard matrix related to the one for
$ q \bar{q} \rightarrow q \bar{q}$, Eq. (\ref{hardqqb}),  by
the crossing transformation $\hat{s} \leftrightarrow \hat{u}$.
The functions $\chi_1$, $\chi_2$ and $\chi_3$ here
are given by
\begin{eqnarray}
\chi_1&=&\frac{\hat{s}^2+\hat{t}^2}{\hat{u}^2}      \nonumber \\
\chi_2&=&N_c \frac{\hat{s}^2}{\hat{t}\hat{u}}-
\frac{\hat{s}^2+\hat{t}^2}{\hat{u}^2}  \nonumber \\
\chi_3&=&\frac{\hat{s}^2+\hat{u}^2}{\hat{t}^2}+
\frac{1}{N_c^2}\frac{\hat{s}^2+\hat{t}^2}{\hat{u}^2}
-\frac{2}{N_c}\frac{\hat{s}^2}{\hat{t}\hat{u}}\,.
\end{eqnarray}
Again, $qq'\rightarrow qq'$ is found by keeping only the
$t$-channel terms.
The lowest-order soft matrix is the same as for $q{\bar
q} \rightarrow q{\bar q}$.

\subsubsection{Hard and soft matrices for $q{\bar q}\rightarrow g g$
at LO}

      In the basis (\ref{ggqbas}), the Born level hard-scattering
matrix is
\begin{equation}
H^{(1)}\left(\hat{t},
\hat{s},\alpha_s \right)= \frac{1}{N_c^2} \,
\frac{\pi \alpha^2_s}{\hat{s}} \, \frac{\hat{t} \hat{u}}{\hat{s}^2}
\left(
\begin{array}{ccc}
\frac{1}{N_c^2}\chi_1    &
\frac{1}{N_c}\chi_1   &
\frac{1}{N_c}\chi_2  \vspace{2mm} \\
       \frac{1}{N_c}\chi_1  & \chi_1 & \chi_2 \vspace{2mm} \\
       \frac{1}{N_c} \chi_2 & \chi_2 & \chi_3
\end{array}
\right) \, ,
\end{equation}
where $\chi_1$, $\chi_2$, and $\chi_3$ are now given by
\begin{eqnarray}
\chi_1 &=& \frac{\hat{t}^2 + \hat{u}^2}{\hat{t}\hat{u}}      \nonumber \\
\chi_2 &=& \left(1+\frac{2\hat{t}}{\hat{s}}\right) \, \chi_1
\nonumber \\
\chi_3 &=& \left(1-\frac{4\hat{t}\hat{u}}{\hat{s}^2}\right) \, \chi_1 \, .
\end{eqnarray}
The zeroth order soft matrix is found to be
\begin{equation}
S^{(0)}= C_F \left(
\begin{array}{ccc} 2N_c^2  & 0 & 0 \\
0 & N_c^2-4 & 0\\
0 & 0 & N_c^2 \end{array} \right) \, .
\label{softqg}
\end{equation}

\subsection{Cross sections}

Given the hard and soft
functions, and the entries of the anomalous
dimension matrices, we have all the ingredients necessary
to calculate the
inclusive cross section (\ref{HSresum}), its differential counterpart
(\ref{lla}), and the ratio $\rho$, Eq.\ (\ref{rhoABdef}), for $p\bar{p}$
scattering in valence approximation.
For a given choice of $\Omega$, such as the patch described above in
terms of azimuthal angles and  rapidities, we find first the functions
$\Gamma^{(ij, \rm f)}$,
$\alpha^{(\rm f)}$, $\beta^{(\rm f)}$ and $\gamma^{(\rm f)}$
for $({\rm f})=(q\bar q \rightarrow  q \bar q),
(q\bar{q} \rightarrow \bar{q} q), (q\bar q \rightarrow g g)$, as in Sec.\ 4.1.
The resulting anomalous dimension matrices are
given by Eqs.\ (\ref{qqbarG}) and (\ref{qqggADM}).  We must
diagonalize these matrices, finding eigenvalues $\lambda_\alpha$, and the
transformation matrices $R^{-1}_{I\beta}$, Eq.\ (\ref{Rdef}), which are used
to take the
hard and soft functions to the diagonal basis space,
as in Eq.\ (\ref{colortransform}).
Finally, the exponents $E_{\alpha\beta}$ that appear
in the cross sections are found from Eq.\ (\ref{Edef}).

\begin{figure}
%\vbox{\vskip 1 true in}
\centerline{\epsfysize=8.8cm \epsffile{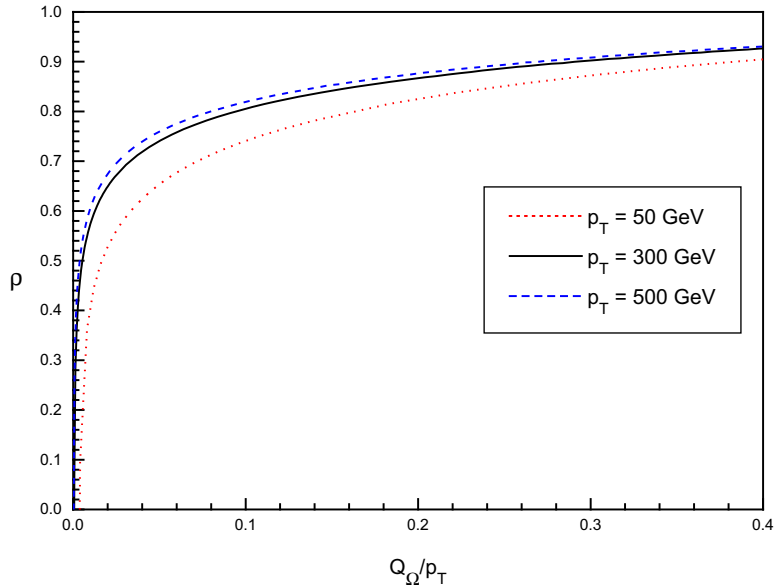}}
\caption{The distributions $\rho(x=Q_\O/p_T)$
for the region $\O$ defined in  Sec.~4.1, with the
observed jet at $\eta=0$ with
$\mu_F = p_T = 50$, 300 and 500 GeV.  This calculation was
carried out in valence quark approximation
for ${\rm p\bar{p}}$ scattering, at ${\protect \sqrt{s}=1.8}$ TeV.}
\label{rhofig}
\end{figure}

As an example, we consider predictions for high-$p_T$
jets at zero hadronic c.m. rapidity ($\eta = 0$)
  in $\rm p\bar p$ scattering
at $\sqrt{s}=1.8$ TeV.   We work in valence quark
approximation, and therefore need only the
exponents for quark-antiquark initial states.  For
$\O$, we take the patch discussed in Sec.\ 4.1, whose exponents
$E_{\alpha\beta}$ were plotted in Figs.\ \ref{Eabfig} and \ref{Eabfigqqgg}.
The resulting ratio $\rho$ is shown in Fig.\ \ref{rhofig}
for $p_T$ = 50, 300 and 500 GeV as a  function of
$x=Q_\O/p_T$, the ratio of the transverse energy flow
to $p_T$, for $0<x<0.4$. (We set $\rho$
to zero below $\Lambda_{\rm QCD}$.) For larger $Q_\O$, we expect
recoil, which we have not  taken into account, to be important.  In
this calculation, we
have chosen the renormalization and factorization scales
equal to $p_T$, and have used CTEQ5L parton distribution functions
\cite{CTEQ}.
The dependence of the ratio $\rho$ on the choice of the factorization
  scale is very weak.

Recall that $\rho(x)$ measures the fraction of events
for which the radiated transverse energy is less than or equal to $xp_T$.
For any choice of $\O$
   the real parts of the exponents $E$, Eq.\ (\ref{Edef}),
are positive.
As a result, the ratio $\rho$ vanishes for $Q_\O=\Lambda_{\rm QCD}$,
and rises rapidly with $x$.
For example, when $p_T=500$ GeV,
roughly 80\% of events have  $Q_\O<50$ GeV for this choice
of $\O$.
In addition, for the choice of
$\O$ given above, ${\rm Re}\, E_{\alpha\beta}$
is almost always less than unity, and
the distribution $d\rho/dQ_\O$, found from
Eqs.\ (\ref{rhoABdef}) and (\ref{lla}), actually diverges
at $Q_\O=\Lambda_{\rm QCD}$.
A similar behavior was found for ``gap" dijet events, with low radiation in
a large region that covers all $0\le \phi < 2\pi$ between
high-$p_T$ jets that are widely-separated in rapidity \cite{StOd}.
In such gap events there is an anomalously  small
eigenvalue, corresponding to a color exchange that approaches pure
singlet in the $t$-channel for $\eta\rightarrow \infty$.  In contrast,
for the region $\O$ considered here, which
is outside the scattering plane
and of limited extent in rapidity,
all the eigenvalues are relatively small.

\begin{figure}
%\vbox{\vskip 1 true in}
\centerline{\epsfysize=8.8cm \epsffile{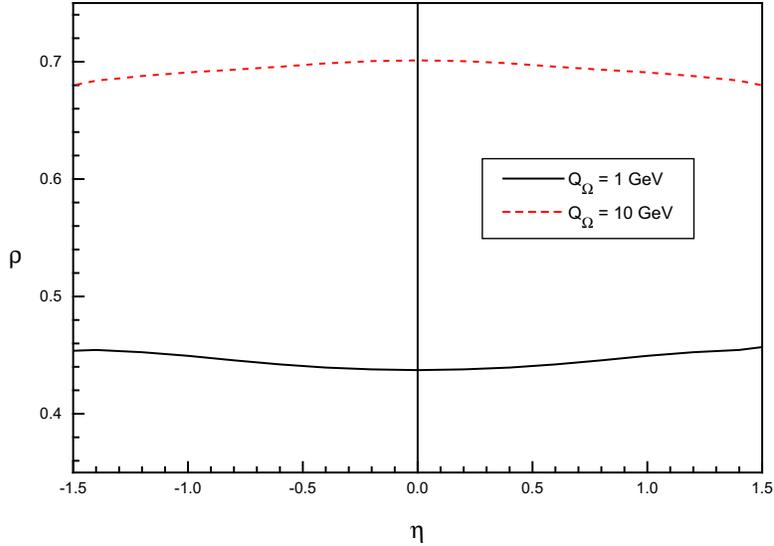}}
\caption{The $\eta$-dependence of $\rho(x)$ for
a fixed region $\O$ defined as in Sec.\ 4.1., for radiated energies
$Q_\O = 1$ GeV and 10 GeV, respectively, and $p_T  = 300$ GeV. }
\label{rhoetafig}
\end{figure}

    The lowest-order expansion
of the differential cross
section $d\hat\sigma^{\rm (f)}/d\hat\eta dQ_\O$,
Eq.\ (\ref{lla}), corresponds to the eikonal approximation
for the NLO correction to the cross section \cite{mw88}.
This cross section is proportional
to the real parts of the  exponents $E_{\alpha\beta}$, Eq.\
(\ref{Edef}),
and is hence proportional to the size of the region $\O$
in $\eta-\phi$ space.
The vanishing of $\rho$ at
low energy flow is an expression of the tendency for
the scattered quarks to radiate, essentially
according to the ``antenna pattern" \cite{coco,cobr,BBEL} of  the
corresponding
classical field.
The divergence of the perturbative one-loop running coupling
is responsible for the vanishing of $\rho(x)$ for $x=\Lambda_{\rm
QCD}/p_T$
rather than at zero, as would be the case for a fixed coupling.
The rapid rise in $\rho(x)$ indicates that most events
generate relatively small amounts of perturbative radiation into  the
patch region.
The differences in the  curves are due to the
running coupling, and become less important as $p_T$ increases.
Finally, in Fig.\ \ref{rhoetafig}, we show  the $\eta$-dependence of
$\rho$ for
two values of $Q_\O$ for the same patch, with  $p_T=300$ GeV.
The rapidity-dependence is mild for this  range in $\eta$ and
choice of $\O$.

\section{Discussion}

The few examples we have given  in the previous section illustrate
the kind of new results on energy flow that can be derived using
the methods outlined in this paper.  In principle, given any fixed
set of jets, or heavy quarks \cite{QQBar}, of  specified
momenta and  directions, we may calculate
the distribution of perturbative
energy flow into any fixed  region $\O$ of rapidity and azimuthal angle
in the eikonal approximation.
After resummation in logarithms of
$Q_\O/p_T$, the distribution
is conveniently described in  terms of  the ratio $\rho(Q_\O/p_T)$,
which specifies the percentage of dijet events for which
a transverse energy of no more
than $Q_\O$ flows  into region $\O$.

The corresponding distribution, $d\rho/d\eta dp_TdQ_\O$ may also be 
interpreted
as a {\it correlation} of transverse energy flows, between the jet
cone and region $\O$, generalizing energy-energy correlations
\cite{BBEL} to ``moving cone"
multijet algorithms, introduced by Giele and Glover \cite{GG}.
These
are examples of Tkachov's  ``jet discriminators" \cite{T}.  In
the formalism of \cite{GG}, the discussion in this paper relates to
a three-cone configuration, in which two cones sample high-$p_T$
jets, while one ``cone", covering $\O$, encounters relatively soft radiation.
We also anticipate that appropriate moments with respect to
transverse energy in $\O$
may be analyzed in the spirit of \cite{eecnp}, to
gain insight into the influence of
nonperturbative corrections for this class of correlations.

In this paper, we have investigated only the
relatively simple case of valence
approximation.
    Beyond this, we hope that the kind of energy flow analysis
     described here will provide an additional tool to study more complex
jet events and production mechanisms at short
distances \cite{EKS}.
Our predictions are at present limited
by our neglect of ``non-global" logarithms, which may be studied
numerically \cite{DasSal01}.  We anticipate that by
specifying restrictions on radiation into the intermediate
region $\bar\O$ more completely, it will be possible to
regain a closed form for contributions of radiation
into $\O$ to certain inclusive cross sections.  These and
related issues are under study.

\subsection*{Acknowledgements}

We thank Anton Chuvakin, Chi Ming Hung,
Gianluca Oderda, and Jack Smith for
many very helpful conversations.
We thank Mrinal Dasgupta and Gavin Salam, for
instructive exchanges, and for communicating their
results prior to publication.
We also benefited greatly from discussions of jets and energy flow
with many colleagues at Snowmass 2001,
including Jon Butterworth, Steve Ellis, Daniel Elvira,
Brenna Flaugher, Walter Giele, Anna Kulesza, and Nikos Varelas.
G.S.\ would like to thank Brookhaven National Laboratory for
its hospitality, and C.F.B.\ thanks the Division of Particles and Fields of
the American Physical Society for support at Snowmass 2001.
This work was supported in part by the National Science
Foundation, grant PHY9722101.

\end{document}